\documentclass[a4paper,twocolumn,superscriptaddress,prx, aps, floatfix ]{revtex4-2}
\usepackage{latexsym,amsmath}
\usepackage{amsbsy}
\usepackage[pdftex]{graphicx}
\usepackage{amssymb}
\usepackage{epstopdf}
\usepackage{verbatim}
\usepackage[usenames]{color}
\usepackage{capt-of} 
\usepackage{float}
\usepackage{ esint }
\usepackage{hyperref}
\usepackage{graphicx}
\usepackage[normalem]{ulem}

\usepackage{xcolor}

\newcommand{\cm}{\langle \widetilde{c}\rangle}

\newcommand{\kl}[1]{\langle k^{(#1)}\rangle}
\newcommand{\km}{\langle \widetilde{k}\rangle}
\begin{document}

	\title{Multiplexity amplifies geometry in networks}
	
	\author{Jasper van der Kolk}
	\email{vanderkolkj@ceu.edu}
	\affiliation{Departament de F\'isica de la Mat\`eria Condensada, Universitat de Barcelona, Mart\'i i Franqu\`es 1, E-08028 Barcelona, Spain}
	\affiliation{Universitat de Barcelona Institute of Complex Systems (UBICS), Barcelona, Spain}
	
	\author{Dmitri Krioukov}
	\affiliation{Department of Physics \& Network Science Institute \&\\ Department of Mathematics \& Department of Electrical and Computer Engineering, Northeastern University, Boston, Massachusetts, USA}
	
	\author{Mari\'an Bogu\~n\'a}
	\email{marian.boguna@ub.edu}
	\affiliation{Departament de F\'isica de la Mat\`eria Condensada, Universitat de Barcelona, Mart\'i i Franqu\`es 1, E-08028 Barcelona, Spain}
	\affiliation{Universitat de Barcelona Institute of Complex Systems (UBICS), Barcelona, Spain}
	
	\author{\\M. \'Angeles \surname{Serrano}}
	\email{marian.serrano@ub.edu}
	\affiliation{Departament de F\'isica de la Mat\`eria Condensada, Universitat de Barcelona, Mart\'i i Franqu\`es 1, E-08028 Barcelona, Spain}
	\affiliation{Universitat de Barcelona Institute of Complex Systems (UBICS), Barcelona, Spain}
	\affiliation{Instituci\'o Catalana de Recerca i Estudis Avan\c{c}ats (ICREA), Passeig Llu\'is Companys 23, E-08010 Barcelona, Spain}
	
	\begin{abstract}
    Many real-world network are multilayer, with nontrivial correlations across layers. Here we show that these correlations amplify geometry in networks. We focus on mutual clustering---a measure of the amount of triangles that are present in all layers among the same triplets of nodes---and find that this clustering is abnormally high in many real-world networks, even when clustering in each individual layer is weak. We explain this unexpected phenomenon using a simple multiplex network model with latent geometry: links that are most congruent with this geometry are the ones that persist across layers, amplifying the cross-layer triangle overlap. This result reveals a different dimension in which multilayer networks are radically distinct from their constituent layers.
	\end{abstract}
	
	\maketitle

	Complex networks are indispensable tools for mapping the intricate relationships between units in many real-world systems. These units are often connected through different types of interactions~\cite{Little2002,Verbrugge1979}, which lead to distinct network topologies. This observation led to the study of multiplex networks, multilayer structures in which nodes are shared across layers, but where connectivity profiles can differ between them~\cite{Lewis2008,Mucha2010,Gomez2013,Szell2010,Morris2012,Cardillo2013,Sole2013,Bianconi2018,Artime2022}. Even though the individual networks that make up the multiplex are, \textit{a priori}, different, the systems these layers represent often exhibit correlations~\cite{Baxter2016,Bianconi2013,Kleineberg2016,Papadopoulos2019}. While several works have studied the structural overlap resulting from these correlations~\cite{Battiston2014,Boccaletti2014,Kim2021,Luu2019,Wang2020,Battiston2017,Dimitrova2020}—that is, the presence of edges between the same pair of nodes in multiple layers of a multiplex— a comprehensive analysis has been lacking.
	
	In this work, we aim to bridge this gap by analyzing the overlap of higher order structures such as triangles. We first survey a large set of real-world multiplexes from different domains and find that while edge overlap is indeed relatively high compared to the uncorrelated baseline~\cite{Bianconi2013}, this effect is even stronger for triangles. We then explain these result using the framework of network geometry~\cite{Boguna2021}, where nodes are assumed to lie in a hidden metric space that determines the likelihood of their connections. The underlying geometry explains the high levels of clustering --a measure of the amount of triangles-- observed in single layer data, even when the coupling of the topology to the latent space is weak~\cite{Serrano2008,vanderKolk2022}. This is a direct consequence of the triangle inequality, which states that two nodes that are close to a third in a metric space must also be close to each other. In real multiplexes, the correlations observed between the latent coordinates of the nodes in the different layers have been described by the geometric multiplex model (GMM)~\cite{Kleineberg2016}, where such correlations can be controlled. Through the introduction of the mutual graph (MG), an auxiliary single layer network made up of those edges that are present in all layers of the multiplex, we study the interplay between geometric correlations in multiplexes and the observed triangle overlap. We find that edges most aligned with the underlying metric space exhibit the greatest overlap. This amplified geometric coupling leads to the formation of many geometrically induced triangles defining a geometric core. This can result in a finite clustering coefficient in the thermodynamic limit, even when this quantity vanishes in all individual layers of the multiplex. Thus, the enhanced geometricity of multiplexes caused by the amplified coupling can help explain the high levels of triangle overlap found in real data. 
	
	We analyzed 15 real-world multiplexes from various fields, whose details can be found in App.~\ref{app:realmultiplexes}. We investigated the overlap properties by pairing layers. That is, if we have a multiplex with $L=4$ layers, we study the overlap of $\binom{4}{2}=6$ unique pairs of layers. 
Then, we extracted the mutually connected component (MCC) of each pair of layers using the algorithm introduced in Ref.~\cite{Hwang2015}. This ensures that, in each layer, there exists a path connecting the analyzed nodes, which is a necessary condition in our geometric model to define a meaningful notion of relative distance between them.
It has also been shown that the MCC is relevant for many applications~\cite{Buldyrev2010,Serrano2015,Son2012}. We will refer to the resulting mutually connected two-layer system as the pairwise MCC. Note that we only analyze these objects if they contain more than 100 nodes because smaller MCCs are prone to finite size effects. Once this procedure has been completed we construct the MG using the same set of nodes as in the pairwise MCC and by keeping only those edges that are present in both layers. Of course, we could define our mutual graph differently. For instance, one could keep edges as long as they were present in either of the two layers. However, our definition of the MG, which corresponds to the intersection of the two layers, is particularly useful for measuring overlap as defined in previous works~\cite{Bianconi2013,Battiston2014}.
	
\begin{figure}
		\centering
		\includegraphics[width=\linewidth]{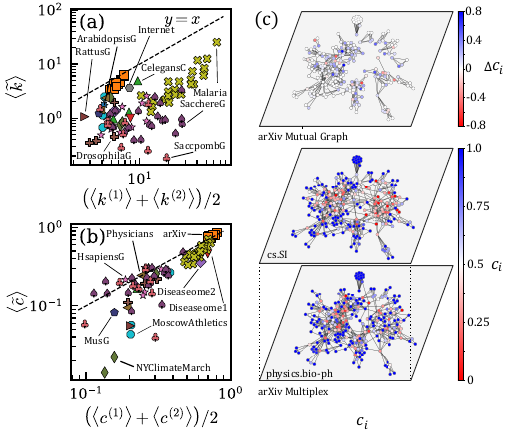}
		\caption{(a) The average mutual degree of the MG versus the mean of the two average degrees of the single layer networks for several real multiplexes. One multiplex can result in several data points as multiplexes with more than two layers are deconstructed into all possible pairwise MCCs. The style of the data points is the same for all MCCs corresponding to the same multiplex, and we only label each type once. For example, MCCs related to the Malaria multiplex are represented by gray hexagons. The black dashed line indicates the diagonal $\langle \widetilde k\rangle = (\kl{1}+\kl{2})/2$. (b) The analogous data for the average local mutual clustering coefficient. The details of the networks shown in this figure can be found in Appendix~\ref{app:realmultiplexes}. (c) An arXiv pairwise MCC and its mutual graph. In the bottom two layers, representing the pairwise MCC, nodes are colored by the local clustering coefficient $c_i$. In the top layer, representing the MG, the nodes are colored by the difference between the local mutual clustering coefficient and the mean of the clustering of the corresponding nodes in the constituent layers $\Delta c_i = \widetilde c_i - (c^{(1)}_i+c^{(2)}_i)/2$. In all layers, nodes with $k_i\leq 1$ are not shown as they do not contribute to the average local clustering coefficient.}
		\label{fig:fig1}
	\end{figure}

First, the average degree $\km$ of the MG, referred hereafter as the average mutual degree, measures the amount of edge overlap between two layers. We note that two sparse, independent Erd\H os-Renyi (ER) network realizations show next to no overlap. There, the probability of a link being present in both layers is $(\langle k\rangle/N)^2$, where $\langle k\rangle$ is the average degree and $N$ is the size of the network. This leads to a graph density in the MG that scales as $\sim N^{-2}$, or an average mutual degree of order $\sim N^{-1}$, resulting in very low values of both quantities in large scale networks. In fact, it was shown in Ref.~\cite{Bianconi2013} that these scalings hold for random graphs with arbitrary degree sequence. However, Fig.~\ref{fig:fig1}a, where we plot the average mutual degree as a function of the mean average degree of the two layers, shows that the amount of edge overlap in real multiplexes is much higher than the null-model would predict. This is in line with past results ~\cite{Bianconi2013,Battiston2014,Boccaletti2014} and with the intuition that there should be correlations between layers leading to similar connectivity patterns. We note, however, that the MG generally has an average mutual degree much lower than the individual layers it is based on. In fact, $\km<1$ for many pairwise MCCs.

For a measure of the amount of overlapping triangles in the pairwise MCCs, we focus on the average local clustering coefficient $\cm$ of the corresponding MG, called hereafter the mutual clustering coefficient. In order for a triangle to be present in the MG, it needs to exist in both layers. Intuitively, this is highly unlikely in the case of independent graphs. In the Supplementary Information (SI)~\cite{supp}, which includes Refs.~\cite{Boguna2004,vanderKolk2022}, we show that the mutual clustering coefficient scales as $\sim N^{-2}$ for sparse ER networks. However, Fig.~\ref{fig:fig1}b tells a different story. Here we plot the average local clustering of the MG against the mean of the two single-layer clustering coefficients. We see that not only does the mutual clustering not tend to zero, it remains of the same order as that of the single layer graphs. In some cases it even exceeds these levels. These results indicate that the correlations present between layers in multiplex networks have an especially strong effect on the triangle overlap.

These results points towards mutual graphs that are sparser than their corresponding individual layers while at the same time showing similar levels of clustering. In Fig.~\ref{fig:fig1}c we visualize this effect for two layers of the arXiv multiplex. We notice that the mutual graph indeed has a reduced amount of edges, even disintegrating into several disconnected components. At the same time, most nodes maintain or increase their local clustering coefficient with respect to the mean of the local clustering coefficient in the constituent graphs ($\Delta c_i = \widetilde{c}_i-(c^{(1)}_i+c^{2}_i)/2 \geq 0$). 
	
Clustering is an essential geometric graph property, linked to the triangle inequality inherent to all metric spaces. It is therefore appropriate to investigate the reported empirical observations through the lens of network geometry~\cite{Boguna2021}. In particular, we adopt the similarity$\times$popularity picture~\cite{Serrano2008,Papadopoulos2012}, in which nodes lie in an underlying similarity metric space modeled as a circle, where nearby vertices are thought to be more similar and therefore more likely to be connected. Simultaneously, some nodes are more popular than others, implying they will form more connections and end up with higher degrees~\cite{Serrano2008}. 
	
In the multiplex setting, this idea is formally implemented through the geometric multiplex model (GMM)~\cite{Kleineberg2016} (see Appendix~\ref{app:detailsGMM} for details). In the GMM, the different single layer networks making up the multiplex are embedded in an underlying similarity space, such that each node is assigned an angular coordinate $\theta^{(l)}_i$ in each layer and pairs of nodes are separated by similarity distances $d^{(l)}_{ij}=R\Delta\theta^{(l)}_{ij}$, where $\Delta\theta^{(l)}_{ij}=|\pi-|\pi-|\theta^{(l)}_i-\theta^{(l)}_j||$. The popularity dimension is incorporated by assigning to each node a hidden degree $\kappa^{(l)}_i$, which are chosen such that the observed degree distribution in a certain layer $l$ has a power law tail $P^{(l)}(k)\sim k^{-\gamma_l}$. The connection probability between two nodes is
\begin{equation}
	p^{(l)}_{ij}\sim (d^{(l)}_{ij})^{-\beta_l}(\kappa^{(l)}_i\kappa^{(l)}_j)^{\max(1,\beta_l)}\label{eq:pijS1}.
\end{equation}
These steps imply that each layer can be seen as a realization of the single-layer geometric $\mathbb{S}^1$-model~\cite{Serrano2008}. In the GMM, the correlations between the angular coordinates across layers are controlled through the parameter $g$. Perfectly correlated multiplexes will have equal coordinates in all layers and correspond to $g=1$. When $g=0$, the layers are uncorrelated. In this case, a node's coordinate in one layer is drawn independently from that node's coordinats in another layer. Similarly, the correlations of hidden degrees are controlled through the parameters $\nu$. The parameter $\beta_l$ controls the geometric coupling strength of the $l$-th layer to its underlying similarity space, determining how strongly the distance between two nodes influences the probability of a connection between them. In the single-layer case, $\beta$ is the tuning parameter of a clustering phase transition~\cite{Serrano2008}. In the strongly geometric regime, with $\beta>1$, clustering is finite. If $\beta\leq1$, corresponding to the weakly geometric regime, it vanishes in the thermodynamic limit. Nevertheless, this decay is very slow, especially in the quasi-geometric regime $\beta_c'<\beta\leq1$~\cite{vanderKolk2022}. Here, the exact value of the transition point $\beta_c'$ depends on the degree distribution.

	\begin{figure*}
		\centering
		\includegraphics[width=\linewidth]{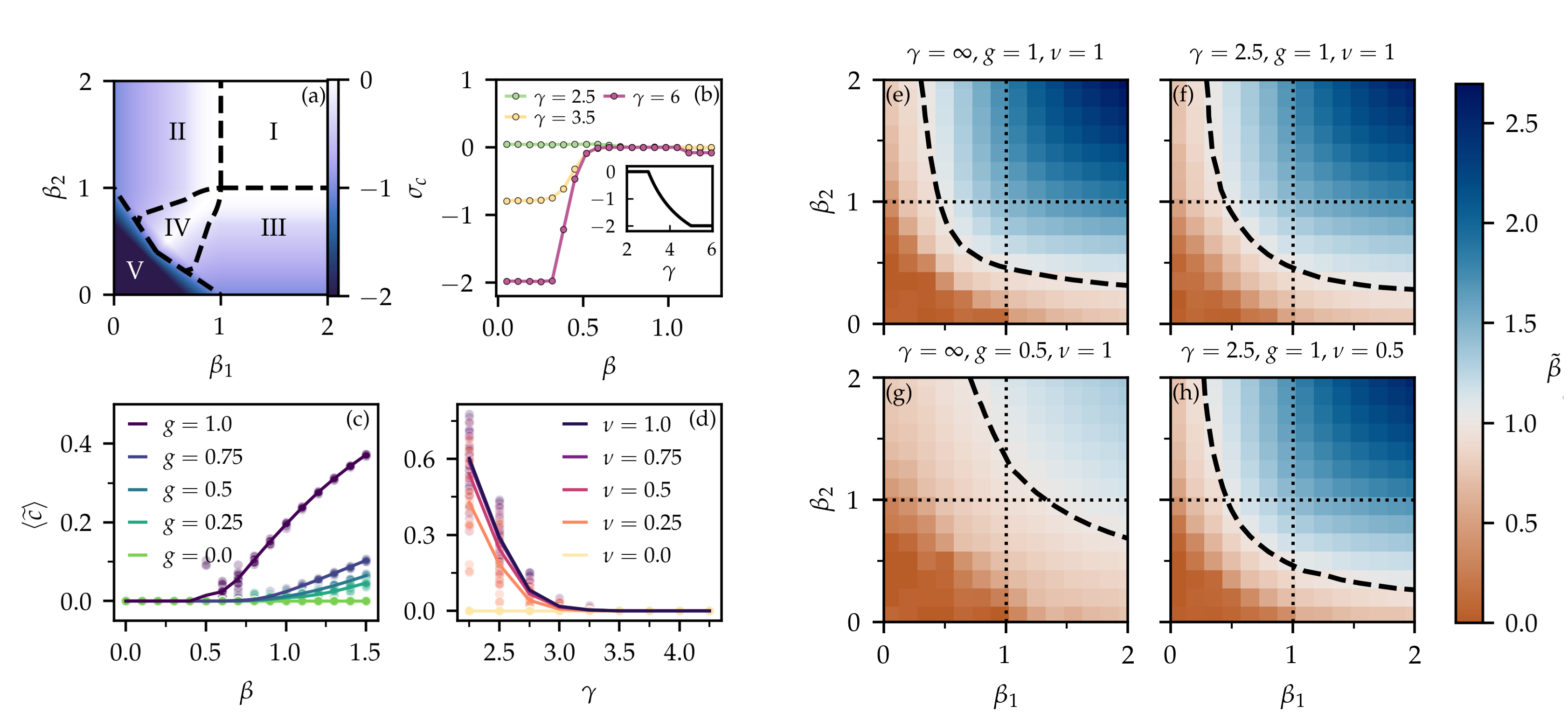}
		\vspace{-8mm}
		\caption{(a) The scaling exponent $\sigma_c$ of the mutual clustering coefficient $\cm \sim N^{\sigma_c}$ as a function of the geometric couplings $\beta_1$ and $\beta_2$ of the constituent graphs for fully correlated layers and homogeneous degree distribution. The black dashed lines define five non-overlapping regions of the parameter space, based on how the mutual clustering coefficient relates to its single layer counterparts. The parametrization of these lines can be found in the SI~\cite{supp} (b) The scaling exponents $\sigma_c$ as a function of $\beta=\beta_1=\beta_2$ for various power-law exponents $\gamma$. Results were obtained by numerical integration of Eq.~S42 in the SI~\cite{supp} for $N\in[10^6,10^9]$.
The inset shows the analytic result (see the SI~\cite{supp} for details) for $\sigma_c$ as a function of $\gamma$ when $\beta=0$. (c) The mutual clustering coefficient as a function of $\beta=\beta_1=\beta_2$ for various correlation strengths $g\in[0,1]$. In all cases the individual layers were generated with $N=32000$, $\gamma=50$, $\langle k\rangle = 20$ and $\nu=1$. (d) The mutual clustering coefficient as a function of the degree distribution exponent $\gamma=\gamma_1=\gamma_2$. Various correlations strengths $\nu\in[0,1]$ are shown. The individual layers use $N=32000$, $\beta=0$, $\langle k\rangle =20$ and $g=1$. (e-h)The relation between the geometric couplings of the MG and that of its constituent graphs, for homogeneous (e,g) and heterogeneous degree distributions (f,h). The black, dotted line corresponds to the effective coupling $\widetilde\beta=1$. Perfect correlations are studied in panels (e,f) whereas weaker correlations are explored in (g,h). For all realizations $N=1500$ and $\kl{1}=\kl{2}=50$.}
		\label{fig:fig2}
	\end{figure*}

	We first investigate if this geometric framework can explain the clustering results obtained in Fig.~\ref{fig:fig1}b. The clustering coefficient of a node $i$ is the probability that a pair of randomly chosen neighbors are neighbors themselves. In Appendix~\ref{app:hiddenvariablemodels} we generalize the results from Ref.~\cite{Boguna2003}, where hidden variable models for single layer networks were studied, to the multiplex setting. This allows us to obtain exact scaling results for the average clustering coefficient in the MG for some limiting cases. 
	
	The first case is that of perfectly correlated layers ($g,\nu=1$) with homogeneous degree distributions. In the SI~\cite{supp} we show analytically that, here, clustering scales with the system size as a power-law $\cm\sim N^{\sigma_c(\beta_1,\beta_2)}$. In Fig.~\ref{fig:fig2}a, we plot the dependence of the exponent $\sigma_c(\beta_1,\beta_2)\equiv \sigma_c$ on the geometric couplings of the two constituent networks. We can distinguish five regions based on whether the mutual clustering decays faster or slower than its single layer counterparts. In region $\text{I}$, $\beta_1,\beta_2>1$ and so both layers are highly clustered. This behavior is carried over to the MG, where clustering remains finite, and so $\sigma_c(\beta_1,\beta_2)=0$. In regions $\text{II}$ and $\text{III}$, one layer is either in the geometric or quasi-geometric regime, whereas the other is weakly geometric. The scaling behavior of the mutual clustering then lies in between that of the individual layers. In region $\text{IV}$, both layers are weakly geometric with $\beta_1,\beta_2\leq 1$. However, the decay of the clustering coefficient is slower than that of the individual layers, implying the presence of significant levels of overlapping triangles. In particular, when $1\geq \beta_1=\beta_2\geq 1/2$, mutual clustering remains finite for all $N$ even though it vanishes in the large size limit of each separate layer. The fact that clustering decays very slowly in these regions is in line with the empirical results in Fig.~\ref{fig:fig1}b. Only when $\beta_1$ and $\beta_2$ are both very small (region $\text{V}$) does clustering decay fast. In fact, the scaling here is equivalent to that of the explicitly non-geometric case where $\beta_1=\beta_2=0$.

	In the SI~\cite{supp} we generalize these results to the $L$-layer case. We show that the region of constant clustering is enhanced for MGs constructed from more layers. In the most extreme case, when the $\beta_l$'s in all $L$ layers are equal, the transition point shifts to $\beta_c=1/L$. This implies that, here, the clustering phase transition vanishes in the large $L$ limit so that clustering in the mutual graph becomes size independent.
	
	Real networks are rarely homogeneous and we, therefore, investigate numerically the effect of degree heterogeneity on the scaling of the clustering coefficient, focusing on the region $\beta_1=\beta_2\equiv\beta$. In Fig.~\ref{fig:fig2}b, $\sigma_c$ is plotted against $\beta$ for $\gamma=2.5$, $\gamma=3.5$ and $\gamma=6$. We observe that the constant clustering for $\beta>1/2$ found in the homogeneous case is also present for heterogeneous networks. In fact, for $2<\gamma<3$ this region is extended all the way down to $\beta=0$. For networks with less pronounced heterogeneity ($\gamma>3$), the clustering does start to decrease for $\beta<1/2$. In the SI~\cite{supp} we verify these results with analytic solutions for the explicitly non-geometric case $\beta=0$. The result of these calculations are shown in the inset of Fig.~\ref{fig:fig2}b. In Figs. \ref{fig:fig2}c,d we explore the effect of imperfect correlations across layers ($g,\nu <1$). In all cases, reduced correlations lead to weaker mutual clustering.
	
	The fact that both degree heterogeneity as well as geometry can lead to high clustering raises the question as to which effect is more important for the real multiplexes described in Fig.~\ref{fig:fig1}. In the SI~\cite{supp} we perform degree preserving randomization~\cite{Cobb2003} on the individual multiplex layers, effectively decoupling them from their underlying metric spaces. We show that in almost all cases, this strongly decreases the mutual clustering, highlighting the importance of geometry. We also note that the decrease is less pronounced, or even absent, in the case of high degree heterogeneity, in line with our findings in Fig.~\ref{fig:fig2}b.
	
	
	We perform a similar analysis for the average mutual degree $\langle \widetilde k\rangle$ in the SI~\cite{supp}. Correlations in both the similarity as well as the popularity dimensions can lead to increased levels of edge overlap. However, the extended regions of constant overlap found for triangles are not observed in the case of edges. Only when both layers are strongly geometric ($\beta_{1,2}>1$) does one observe a constant average mutual degree. This is once again in line with our observations in real-world multiplexes, where edge-overlap is relatively low in comparison to that of triangles. It also highlights the fact that high mutual clustering can exist even when edge overlap is low.

	Thus far, we have shown that the GMM can reproduce our findings for real data. However, we still lack an explanation as to why the MG has such surprisingly high levels of clustering. In the single layer case, a higher geometric coupling implies higher levels of clustering due to a larger influence of the triangle inequality present in the underlying metric space. It is therefore pertinent to study the effective geometric coupling of the MG. 
	
	To this end we employ the tool \textit{Mercator}~\cite{GarciaPerez2019,vanderKolk2024a}, which uses a combination of machine learning and maximum likelihood techniques to faithfully embed networks into their hidden popularity$\times$similarity spaces. It has been shown that artificial networks generated using this embedding reproduce important features of the original network, such as the degree distribution and the clustering spectrum. In this paper we specifically leverage the tool’s estimation of the parameter $\beta$ from the observed clustering coefficient. 
	
	\begin{figure}
		\centering
		\includegraphics[width=\linewidth]{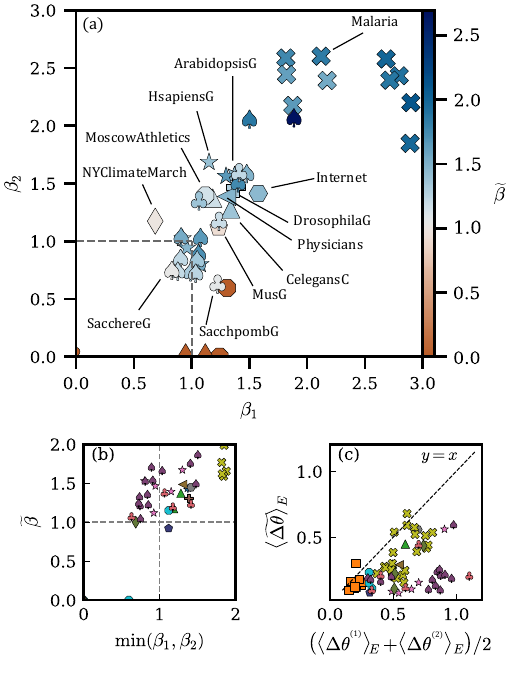}
		\vspace{-7mm}
		\caption{(a) The relation between the geometric couplings of the mutual graph and constituent single layer networks for several real multiplexes. White color corresponds to the transition point $\widetilde{\beta}=1$. (b) The effective coupling $\widetilde \beta$ is plotted against the smallest of the geometric couplings of the underlying layers. (c) The relationship between the average angular distance of connected nodes in the mutual graph and the mean of the average angular distances in the two corresponding single-layer MCCs.}
		\label{fig:realnets}
	\end{figure}
	
	We first embedded several artificial mutual graphs constructed from pairs of $\mathbb{S}^1$ networks at varying $\beta_1$ and $\beta_2$ to obtain their  effective geometric coupling $\widetilde\beta$. In order to obtain a fully connected MG, we set the average degrees of the individual layers to be relatively large ($\kl{1}=\kl{2}=50$). The results of this procedure are shown in Figs.~\ref{fig:fig2}e-h. There, the dotted line indicates the contour $\widetilde{\beta}=1$, which marks the transition point between the strongly and weakly geometric regimes. In panel (e) we focus on homogeneous degree distributions and perfect correlations. Notably, the supercritical regime $\widetilde\beta>1$ extends deep into the region where one or two of the constituent layers is weakly geometric. This occurs because the $\mathbb{S}^1$ connection probability is a decreasing function of the distance, meaning that nodes within individual layers predominantly form short-range connections. It is therefore much more likely that these are present in both layers. These short links are more closely aligned with the underlying metric space, which strengthens the effective geometric coupling, $\widetilde\beta$, and explains how weakly geometric layers can produce strongly geometric mutual graphs. This effect is strongest when the geometric couplings of the individual layers are equal. In the Supplementary Material~\cite{supp} we show that, in this case, the geometric coupling in the GM is given by $\widetilde\beta\simeq\beta_1+\beta_2\equiv 2\beta$. The enhanced supercritical regime is robust to variations in the degree distribution, as evidenced 
in panel (f), which incorporates degree heterogeneity. Panels (g) and (h) show that while reducing correlations in the similarity dimension (e.g. setting $g=0.5$) leads to more weakly geometric mutual graphs, variations in the correlation strength within the popularity dimension have little effect.

Finally, we performed the same analysis on the real-world multiplexes. We embedded the individual layers of their pairwise MCCs as well as the corresponding MG. Note that the MG for empirical networks generally has a low average mutual degree, and might not be fully connected. In this case we only embed the giant connected component (GCC) if this network has more than fifty nodes.
The behavior in Fig.~\ref{fig:realnets}a is in agreement with the model as the inferred effective coupling $\widetilde{\beta}$ is mostly high. In Fig.~\ref{fig:realnets}b we see that generally $\widetilde{\beta}>1$, even when one of the two layers is weakly geometric. In Fig.~\ref{fig:realnets}c we show that it is mostly the short, geometric edges that exist in both layers of the multiplex, leading to a shorter average distance $\langle \Delta\theta\rangle_E$ between the connected nodes in the mutual graph than in the individual layers. Of course, the increase in the geometricity of the empirical MGs is attenuated by the fact that real multiplexes do not have perfect correlations between the layers~\cite{Kleineberg2016,Larremore2013}. However, the enhanced geometry in these real multiplexes, where the mutual graph consists of mostly short, geometric edges, helps explain the high levels of triangle overlap reported in Fig.~\ref{fig:fig1}b.

In this paper, we have shown that edge and triangle overlap are higher than expected in many real multiplexes, with the effect being especially stark for triangles. We analyzed these results through the lens of network geometry, showing that the geometric multiplex model explains real-world observations. The edges that are most congruent with the underlying metric space are the ones present in all layers of the multiplex. This leads to an enhanced effective geometric coupling in the mutual graph, which is made up of the edges present in all layers. We confirm this effect in both synthetic and real-world multiplexes. 

We expect this enhanced geometry, and the resulting triangle overlap, to have significant implications for dynamical processes unfolding on such multiplex systems. Geometric correlations are known to influence a variety of dynamical processes, including navigability using protocols such as Greedy Routing (GR)~\cite{Boguna2009b} and the formation of Turing patterns in reaction-diffusion systems~\cite{vanderKolk2023}, as well as mitigating the extreme vulnerability of multiplex networks to targeted attacks~\cite{Kleineberg2017b}. Consequently, by controlling the geometric amplification and with it the levels of clustering in the multiplex, it is possible to influence transition thresholds, communicability, pattern formation, and network resilience. Furthermore, it was shown in Ref~\cite{Wang2020} that overlapping edges have a strong effect on spreading dynamics on multiplexes. These avenues will be explored in future work, within a broader context that investigates the interplay between dynamical processes and geometric networks. Finally, we note that, much like multiplex percolation~\cite{Cellai2013,Hackett2016} or the rich phase space of multilayer superconductor networks~\cite{Bonamassa2023}, the emergence of multilayer-enhanced geometry is a genuinely new phenomenon—one that has no analogue in the single-layer setting. 

\begin{acknowledgments}
J.K.\ acknowledges support from the Ministry of Universities of Spain in the form of the FPU predoctoral contract as well as well as from the AccelNet-MultiNet program, a project of the National Science Foundation (Award \#1927425 and \#1927418). D.K.\ acknowledges NSF Grant Nos.\ IIS-1741355 and CCF-2311160. M.B.\ and M.A.S.\ acknowledge grant TED2021-129791B-I00 funded by MICIU/AEI/10.13039/501100011033 and the ``European Union NextGenerationEU/PRTR''; grant PID2022-137505NB-C22 funded by MICIU/AEI/10.13039/501100011033 and by ERDF/EU; Generalitat de Catalunya grant number 2021SGR00856. M.B.\ acknowledges the ICREA Academia award, funded by the Generalitat de Catalunya.
\end{acknowledgments}

\section*{Data Availability}
The sources of the real data analyzed in this paper can be found in App.~\ref{app:realmultiplexes}. The corresponding results in tabular form are provided in the Supplementary Information~\cite{supp}. The implementation and outputs of the numerical simulations performed in this study are available from the authors upon reasonable request.

	\pagebreak

	\appendix
	
	\section{Details real networks}\label{app:realmultiplexes}
	The following is a summary of the details of the real multiplexes used in this papers. Further details can be found in the SI~\cite{supp}. 
	\begin{itemize}
		\item \textbf{ArabidopsisG}, \textbf{DrosophilaG}, \textbf{HsapiensG}, \textbf{MusG}, \textbf{RattusG}, \textbf{SacchcereG}, \textbf{SacchpombG}~\cite{DeDomenico2015}: Genetic multiplex networks for various organisms. The layers represent different types of genetic interactions. 
		
		\item \textbf{arXiv}~\cite{DeDomenico2015b}: A coauthorship network among researchers who posted preprints to arXiv, where all papers up to May 2014 are taken into account. Layers represent one-mode projections from underlying author-paper bipartite networks for different subfields.
		
		\item \textbf{CelegansC}~\cite{Chen2006}: A connectome of the flatworm Caenorhabditis elegans. The different layers represent different synaptic junctions.
		
		\item \textbf{Diseasome1}, \textbf{Diseasome2}~\cite{Halu2019}: Two multiplex networks of human disease. The first network is based on data from the Genome-wide Association Study (GWAS) whereas the second uses the Online Mendalian Inheritance in Man (OMIM) catalog. Layers represent one-mode projections from underlying bipartite networks based on phenotype and genotype.

		\item \textbf{Physicians}~\cite{Coleman1957}: A sociometric multiplex investigating the adoption of a new drug, tetracycline, by a group of physicians. Layers were generated with results from different questions about the relationships between physicians.
		
		\item \textbf{MoscowAthletics}, \textbf{NYClimateMarch}~\cite{Omodei2015}: Multiplex networks representing activity among users on Twitter during the 2013 world championships in athletics held in Moscow and the 2014 people's climate march held in New York. Different layers represent different types of twitter interactions.
		
		\item \textbf{Internet}~\cite{Kleineberg2016}: The multiplex of the internet at the level of autonomous systems (AS). These AS nodes are parts of the internet infrastructure administered by a single company or organization. Two nodes are connected if they interchange information packets. Different layers represent different types of destination addresses.
		
		\item \textbf{Malaria}~\cite{Larremore2013}: The multiplex of recombinant antigen genes from the human Malaria parasite Plasmodium falciparum, where nodes are var genes encoding for an antigen protein expressed on the surface of the infected red blood cell. These var genes are capable of interchanging bits of genetic information, thus creating a vast amount of slightly different genes that in turn create highly varying antigens that are hard for the human immune system to detect. Two nodes are connected if they share a substring of nucleotides of significant length, indicating that an interchange of genetic material between the two nodes has taken place. The different layers represent different highly variable regions (HRVs).
	\end{itemize}
	
	\section{Geometric Multiplex Model}\label{app:detailsGMM}
	In this section we explain in detail how one can generate geometric multiplexes with tunable correlation strengths using the Geometric Multiplex Model (GMM). This multiplex variant of the $\mathbb{S}^1$-model starts by drawing angular coordinates from the uniform distribution $\mathcal{U}(0,2\pi)$ and hidden degrees from the Pareto distribution $\rho(\kappa)\propto \kappa^{-\gamma}$, where $\gamma$ encodes the heterogeneity of the distribution. These are then the assigned hidden variables of the first layer in the multiplex. 
	
	The similarity coordinates of the second layer are then drawn with respect to the first such that the correlation between them can be tuned. Take a node $i$ in layer $l=1$ with angular coordinate $\theta_i$. We then want to draw a new coordinate from a truncated Gaussian centered around this coordinate:
	\begin{equation}
		\theta_i^{(2)} = \mod\left(\theta_i^{(l)}+\frac{2\pi l_i}{N},2\pi\right),
	\end{equation}
	where $l_i\in[-N/2,N/2]$ is the arc-distance from the original coordinate along the circle with radius $R=N/(2\pi)$. It is drawn from the truncated Gaussian with probability density function
	\begin{equation}
		f_\sigma(l) = \frac{1}{\sigma}\frac{\phi\left(\frac{l}{\sigma}\right)}{\Phi\left(\frac{N}{2\sigma}\right)-\Phi\left(-\frac{N}{2\sigma}\right)},
	\end{equation}
	where $\phi(x)=\frac{1}{\sqrt{2\pi}}e^{-\frac{1}{2}x^2}$ is the probability density function of a standard normal distribution and $\Phi(x)$ is its cumulative distribution function $\Phi(x)=\frac{1}{2}(1+\text{erf}(x/\sqrt{2}))$. The scaling factor $\sigma$, which is related to the variance of the truncated Gaussian, is defined as
	\begin{equation}
		\sigma = \sigma_0\left(\frac{1}{g}-1\right),
	\end{equation}
	where $\sigma_0=\min(100,N/(4\pi))$. It is clear that $g\in[0,1]$ is the parameter that tunes the correlation between the two layers. When $g\rightarrow0$, $\sigma\rightarrow\infty$ and the Gaussian becomes flat. The coordinate of node $i$ in layer $l=2$ is uniformly sampled, irrespective of its location in the first layer. The correlation between the two layers is therefore zero. On the other extreme, when $g=1$, $\sigma$ vanishes, which implies that the Gaussian becomes a Dirac delta, leading to $\theta_i^{(2)}=\theta_i^{(1)}$ for all $i$, and therefore perfectly correlated coordinates between the layers. 
	
	In the popularity dimension we have the added constraint that the marginal distribution of hidden degrees in the second layer should still be Pareto with some average degree $\langle k^{(2)}\rangle$ and exponent $\gamma_2$, both of which might or might not be different to the ones in the first layer. 
	
	In the original publication Ref.~\cite{Kleineberg2016}, it was shown that this is achieved if the hidden degrees in layer $l=2$ are drawn from the following cumulative distribution function:
	\begin{alignat}{6}
		F_\nu&\left(\kappa_i^{(2)}|\kappa_i^{(1)}\right) = \exp\left(-\left(\varphi_1^{1/(1-\nu)}+\varphi_2^{1/(1-\nu)}\right)^{1-\nu}\right)\notag\\[3mm]
		&\times \left(\varphi_1^{1/(1-\nu)}+\varphi_2^{1/(1-\nu)}\right)^{-\nu}\frac{\varphi_1^{\nu/(1-\nu)}\kappa_0^{(1)}\left(\kappa_i^{(1)}\right)^\gamma_1}{\kappa_0^{(1)}\left(\kappa_i^{(1)}\right)^\gamma_1-\kappa_0^{(1)}\kappa_i^{(1)}}\label{eq:Fvkappa},
	\end{alignat}
	where 
	\begin{equation}
		\varphi_l = -\ln\left(1-\left(\frac{\kappa_0^{(l)}}{\kappa_i^{(l)}}\right)^{\gamma_l-1}\right).
	\end{equation}
	It is now the parameter $\nu\in[0,1]$ that sets the strength of the correlation between the two layers. When $\nu=0$ it can be shown that Eq.~\eqref{eq:Fvkappa} reduces to $F_0(\kappa_i^{(2)})=1-\kappa_i^{(2)}\left(\kappa_0^{(2)}\right)^{\gamma_2-1}$, which is just the Pareto cumulative density function, and does not depend on $\kappa_i^{(1)}$. The two layers are thus independent. When $\nu=1$, we see that Eq.~\eqref{eq:Fvkappa} becomes a step function, leading to $\kappa_i^{(2)}=\kappa_0^{(2)}\left(\kappa_i^{(1)}/\kappa_0^{(1)}\right)^{(1-\gamma_1)/(1-\gamma_2)}$. Here, the correlation between the two layers is perfect; the largest hidden degree in layer $l=1$ is related to the largest hidden degree in layer $l=2$ etc.
	
	If more layers are present this process is continued, generating the hidden variables in layer $l$ on the basis of those in layer $l-1$. In principle one could choose to vary $g$ and $\nu$ parameters, leading to some pairs of layers that are more strongly correlated than others. In this work we choose them identical between all layers. 
	
	When all hidden variables are assigned, the nodes in the different layers are connected using connection probability \eqref{eq:pijS1}, where each layer can have a distinct $\hat\mu_l$ and $\beta_l$. The process of connecting the nodes introduces no new correlation into the system; each edge is placed independently. 
	
	\section{Multiplex Hidden Variable models}\label{app:hiddenvariablemodels}
	In this section we extend the hidden variable framework developed in Ref.~\cite{Boguna2003} to multiplex networks. We assume that each node has a $L$-dimensional vector $\boldsymbol{h}$ associated to it, where the entries of the vector represent the different layers. For the two-layer, homogeneous GMM this means that $\boldsymbol{h} = \{\theta^{(1)},\theta^{(2)}\}$. In the following, we assume the entries to be scalars for notational simplicity. However, the results are trivially extendable to higher dimensional entries, required for example in the case of the heterogeneous GMM, where also hidden degrees are present.
	
	\subsection{The average mutual degree}\label{app:hiddenvariablemodels:k}
	Take two nodes with associated hidden variable vectors $\boldsymbol{h}$ and $\boldsymbol{h'}$. The probability that they are connected, here denoted by $\boldsymbol{h}\sim\boldsymbol{h'}$, is given by the mutual connection probability 
	\begin{equation}
		\widetilde{p}(\boldsymbol{h},\boldsymbol{h'}) \equiv P (\boldsymbol{h}\sim \boldsymbol{h'}) =\prod_{l=1}^{L} p_l(h^{(l)},h'^{(l)})\label{eq:hsimh'},
	\end{equation}
	where we have used the fact that the edge placement is uncorrelated. 
	
	To obtain the total amount of edges, we marginalize over the hidden variables $\boldsymbol{h},\boldsymbol{h'}$ and multiply by $\binom{N}{2}$ as all nodes are, \textit{a priori}, identical. To perform this marginalization we need the probability density function of the hidden variable vector $\rho(\boldsymbol{h})$. In line with the GMM, we assume that the assignment of hidden variables in the different layers is Markovian; the hidden variables of layer $l+1$ only depend on layer $l$. This allows us to write the desired probability density function as 
	\begin{equation}
		\rho(\boldsymbol{h}) = \rho(h^{(1)})\prod_{l=2}^L \rho\left(h^{(l)}|h^{(l-1)}\right)
	\end{equation}
	Finally, we obtain for the expected amount of links the following expression
	\begin{equation}
		\langle \widetilde{M} \rangle = \binom{N}{2} \iint d^L\boldsymbol{h}d^L\boldsymbol{h'}\rho(\boldsymbol{h})\rho(\boldsymbol{h'})\widetilde{p}(\boldsymbol{h},\boldsymbol{h'})\label{eq:M_generalh},
	\end{equation}
	where we have introduced the $L$ dimension integral measure $d^L \boldsymbol{h} = \prod_{l=1}^{L}dh^{(l)}$. 
	%
	%
	%
	
	\subsection{The average local clustering coefficient}\label{app:hiddenvariablemodels:c}
	The average local clustering coefficient is defined as the probability that two neighbors of a node are also connected to each other. Say we have a node with hidden variable vector $\boldsymbol{h}$ connected to two nodes with $\boldsymbol{h'}$ and $\boldsymbol{h''}$, respectively. Given this situation, the probability of the triangle being closed is given by Eq.~\eqref{eq:hsimh'}, the mutual connection probability. 
	
	The first step in deriving the average local clustering coefficient is finding the expected clustering coefficient of a node with hidden variable vector $\boldsymbol{h}$. To this end we need to marginalize the mutual connection probability over $\boldsymbol{h'}$ and $\boldsymbol{h''}$, which requires the probability that a node with hidden variable $\boldsymbol{h}$ is connected to a node with $\boldsymbol{h'}$ (and equivalently to a node with $\boldsymbol{h''}$). In Ref.~\cite{Boguna2003} this was derived to be
	\begin{equation}
		P (\boldsymbol{h}\sim\boldsymbol{h'}|\boldsymbol{h}) = \frac{\rho(\boldsymbol{h'})\widetilde p(\boldsymbol{h},\boldsymbol{h'})}{N ^{-1}\mathbb{E}(k|\boldsymbol{h})},
	\end{equation}
    the probability that a randomly chosen node has hidden variables $\boldsymbol{h'}$ times the probability that the node $\boldsymbol{h}$ is connected to it, where normalization has been taking into account. Marginalizing over the hidden variables of the neighbors we obtain 
	\begin{equation}
		\mathbb{E}(c|\boldsymbol{h}) = \frac{\iint d^L\boldsymbol{h'}d^L\boldsymbol{h''}\rho(\boldsymbol{h'})\rho(\boldsymbol{h''})\widetilde p(\boldsymbol{h},\boldsymbol{h'})\widetilde p(\boldsymbol{h},\boldsymbol{h''})\widetilde p(\boldsymbol{h'},\boldsymbol{h''})}{N^{-2}\mathbb{E}(k|\boldsymbol{h})^2}\label{eq:c_generalh}.
	\end{equation}
	Finally, marginalizing over $\boldsymbol{h}$ we obtain the average local clustering coefficient
	\begin{equation}
		\overline{c}=\int d^L \boldsymbol{h}\, \mathbb{E}(c|\boldsymbol{h})\label{eq:c_marginalized_general}.
	\end{equation}


	%

\end{document}


\title{Supplemental Material for \\Multiplexity amplifies geometry in networks}

	\author{Jasper van der Kolk}
\email{vanderkolkj@ceu.edu}
\affiliation{Departament de F\'isica de la Mat\`eria Condensada, Universitat de Barcelona, Mart\'i i Franqu\`es 1, E-08028 Barcelona, Spain}
\affiliation{Universitat de Barcelona Institute of Complex Systems (UBICS), Barcelona, Spain}

\author{Dmitri Krioukov}
\affiliation{Department of Physics \& Network Science Institute \&\\ Department of Mathematics \& Department of Electrical and Computer Engineering, Northeastern University, Boston, Massachusetts, USA}

\author{Mari\'an Bogu\~n\'a}
\email{marian.boguna@ub.edu}
\affiliation{Departament de F\'isica de la Mat\`eria Condensada, Universitat de Barcelona, Mart\'i i Franqu\`es 1, E-08028 Barcelona, Spain}
\affiliation{Universitat de Barcelona Institute of Complex Systems (UBICS), Barcelona, Spain}

\author{\\M. \'Angeles \surname{Serrano}}
\email{marian.serrano@ub.edu}
\affiliation{Departament de F\'isica de la Mat\`eria Condensada, Universitat de Barcelona, Mart\'i i Franqu\`es 1, E-08028 Barcelona, Spain}
\affiliation{Universitat de Barcelona Institute of Complex Systems (UBICS), Barcelona, Spain}
\affiliation{Instituci\'o Catalana de Recerca i Estudis Avan\c{c}ats (ICREA), Passeig Llu\'is Companys 23, E-08010 Barcelona, Spain}

\maketitle

\tableofcontents

\section{Real Networks}

In this section we provide the details of the real networks presented in the main paper. The data used to generate Fig.~1 in the main text is given in Tabs.~\ref{tab:s1} and ~\ref{tab:s2}. In Tab.~\ref{tab:s3} and ~\ref{tab:s4} we provide further insight into the relationships between the different layers through the Spearman's rank correlation coefficient $r_s$. We investigate the correlations in the degree sequences ($\boldsymbol{k}$) between the different graphs as well the sequences of local clustering coefficients ($\boldsymbol{c}$). The data used to generate Fig.~3 in the main text is given in Tab.~\ref{tab:s5}. Note that there are less entries in the last table as we only show networks where $\widetilde{N}_{GCC}>50$. 

\begin{itemize}
	\item \textbf{ArabidopsisG}, \textbf{DrosophilaG}, \textbf{HsapiensG}, \textbf{MusG}, \textbf{RattusG}, \textbf{SacchcereG}, \textbf{SacchpombG}~\cite{DeDomenico2015}: Genetic multiplex networks for various organisms: Rockcress (Arabidopsis thaliana), fruitfly (Drosophila melanogaster), human (Homo sapiens), mouse (Mus musculus) rat (Rattus norvegicus), baker's yeast (Saccharomyces cerevisiae) and fission yeast (Schizosaccharomyces pombe). The layers represent different types of genetic interactions:
	\begin{enumerate}
		\item Direct interaction
		\item Physical association
		\item Additive genetic interaction defined by inequality
		\item Suppressive genetic interaction defined by inequality
		\item Synthetic genetic interaction defined by inequality
		\item Association
		\item Colocalization
	\end{enumerate}
	The original network was directed, where an edge $(i,j)$ indicates gene $i$ interacting with gene $j$. For the purposes of this article we work with the undirected projection. 
	
	\item \textbf{arXiv}~\cite{DeDomenico2015b}: A coauthorship network among researchers who posted preprints to arXiv, where all papers up to May 2014 are taken into account. Layers represent one-mode projections from underlying author-paper bipartite networks for different subfields:
	\begin{enumerate}
		\item Physics and Socieyt (physics.soc-ph)
		\item Data Analysis, Statistics (physics.data-an)
		\item Biological Physics (physics.bio-ph)
		\item Mathematical Physics (math-ph)
		\item Optimization and Control (math.OC)
		\item Disordered Systems and Neural Networks (cond-mat.dis-nn)
		\item Statistical Mechanics (cond-mat.stat-mech)
		\item Molecular Networks (q-bio.MN)
		\item Quantitative Biology (q-bio)
		\item Biomolecules (q-bio.BM)
		\item Adaptation and Self-Organizing Systems (nlin.AO)
		\item Social and Information Networks (cs.SI)
		\item Computer Vision and Pattern Recognition (cs.CV).
	\end{enumerate}
	
	\item \textbf{CelegansC}~\cite{Chen2006}: A connectome of the flatworm Caenorhabditis elegans. The different layers represent different synaptic junctions:
	\begin{enumerate}
		\item Electric (``ElectrJ")
		\item Chemical Monadic (``MonoSyn")
		\item Chemical Polyadic (``PolySyn").
	\end{enumerate}
	
	\item \textbf{Diseasome1}, \textbf{Diseasome2}~\cite{Halu2019}: Two multiplex networks of human disease. The first network is based on data from the Genome-wide Association Study (GWAS) whereas the second uses the Online Mendalian Inheritance in Man (OMIM) catalog. Layers represent one-mode projections from underlying bipartite networks based on
	\begin{enumerate}
		\item Phenotype (shared symptoms)
		\item Genotype (shared genes)
	\end{enumerate}

	\item \textbf{Physicians}~\cite{Coleman1957}: A sociometric multiplex investigating the adoption of a new drug, tetracycline, by a group of physicians. Layers were generated with results from the following questions:
	\begin{enumerate}
		\item ``When you need information or advice about questions of therapy where do you usually turn?"
		\item ``And who are the three or four physicians with whom you most often find yourself discussing cases or therapy in the course of an ordinary week -- last week for instance?"
		\item ``Would you tell me the first names of your three friends whom you see most often socially?"
	\end{enumerate} 
	
	\item \textbf{MoscowAthletics}, \textbf{NYClimateMarch}~\cite{Omodei2015}: Multiplex networks representing activity among users on Twitter during the 2013 world championships in athletics held in Moscow and the 2014 people's climate march held in New York. The different layers represent
	\begin{enumerate}
		\item Retweets
		\item Mentions
		\item Replies
	\end{enumerate}
	
	\item \textbf{Internet}~\cite{Kleineberg2016}: The multiplex of the internet at the level of autonomous systems (AS). These AS nodes are parts of the internet infrastructure administered by a single company or organization. Two nodes are connected if they interchange information packets. Different layers represent different types of destination addresses:
	\begin{enumerate}
		\item IPv4
		\item IPv6
	\end{enumerate}
	
	\item {\color{black}\textbf{Malaria}~\cite{Larremore2013}: The multiplex of recombinant antigen genes from the human Malaria parasite Plasmodium falciparum, where nodes are var genes encoding for an antigen protein expressed on the surface of the infected red blood cell. These var genes are capable of interchanging bits of genetic information, thus creating a vast amount of slightly different genes that in turn create highly varying antigens that are hard for the human immune system to detect. Two nodes are connected if they share a substring of nucleotides of significant length, indicating that an interchange of genetic material between the two nodes has taken place. The different layers represent different highly variable regions (HRVs). 
	\begin{enumerate}
		\item HRV\_1
		\item HRV\_2
		\item HRV\_3
		\item HRV\_4
		\item HRV\_5
		\item HRV\_6
		\item HRV\_7
	\end{enumerate}}
	
\end{itemize}

\begin{table*}[t]
	\centering
	\caption{\label{tab:s1} Properties of the pairwise mutually connected components (MCC) for several real multiplexes. Column 2 and 3 provide the pair of layers used to construct the pairwise MCC. Column 4 provides the size of the MCC ($N_{MCC}$), which is equal to the size of the mutual graph ($\widetilde N$). Columns 5,6,11 represent the average degrees ($\langle k\rangle$), columns 7,8,11 provide the maximal degree in the network ($k_c)$ and finally, columns 9,10,12 give the average local clustering coefficients ($\langle c\rangle$).}  
	\scriptsize
	\begin{ruledtabular}
		\begin{tabular}{lll|ccccccc|ccc}
			\hline
			Multiplex &  $l$ &  $m$ & $N_{MCC}=\widetilde{N}$ &  $\langle k^{(l)}\rangle$ &  $\langle k^{(m)}\rangle$ & $k_{c}^{(l)}$  & $k_{c}^{(m)}$  & $\langle c^{(l)}\rangle$  & $\langle c^{(m)}\rangle$  &  $\langle \widetilde k\rangle$ &  $\widetilde{k}_{c}$ &  $\langle \widetilde c\rangle$ \\ \hline
			ArabidopsisG & 1 & 2 & 442 & 4.8 & 3.6 & 44 & 29 & 0.26 & 0.28 & 1.8 & 22 & 0.22 \\ 
			arXiv & 3 & 5 & 100 & 5.0 & 4.6 & 20 & 17 & 0.72 & 0.75 & 3.5 & 16 & 0.74 \\ 
			arXiv & 5 & 6 & 182 & 4.7 & 5.2 & 30 & 34 & 0.75 & 0.72 & 3.6 & 29 & 0.80 \\ 
			arXiv & 2 & 6 & 916 & 6.8 & 5.7 & 74 & 59 & 0.66 & 0.72 & 4.7 & 56 & 0.76 \\ 
			arXiv & 6 & 12 & 521 & 5.6 & 5.7 & 46 & 44 & 0.72 & 0.68 & 3.8 & 34 & 0.78 \\ 
			arXiv & 5 & 12 & 310 & 5.3 & 5.9 & 33 & 29 & 0.79 & 0.72 & 4.6 & 26 & 0.81 \\ 
			arXiv & 2 & 7 & 210 & 7.4 & 6.3 & 36 & 27 & 0.77 & 0.84 & 6.0 & 27 & 0.86 \\ 
			arXiv & 3 & 6 & 790 & 5.2 & 5.4 & 42 & 44 & 0.72 & 0.70 & 4.4 & 38 & 0.75 \\ 
			arXiv & 2 & 5 & 506 & 6.9 & 5.5 & 73 & 63 & 0.73 & 0.81 & 5.3 & 63 & 0.82 \\ 
			arXiv & 2 & 3 & 564 & 6.5 & 5.5 & 47 & 42 & 0.67 & 0.74 & 4.2 & 38 & 0.78 \\ 
			arXiv & 3 & 12 & 297 & 5.5 & 5.3 & 33 & 29 & 0.74 & 0.67 & 3.5 & 22 & 0.80 \\ 
			arXiv & 2 & 12 & 2252 & 7.1 & 6.5 & 80 & 60 & 0.76 & 0.79 & 6.3 & 60 & 0.80 \\ 
			CelegansC & 2 & 3 & 257 & 6.9 & 12.1 & 46 & 77 & 0.21 & 0.31 & 4.9 & 41 & 0.20 \\ 
			CelegansC & 1 & 3 & 247 & 4.1 & 11.3 & 40 & 78 & 0.24 & 0.30 & 1.3 & 28 & 0.31 \\ 
			CelegansC & 1 & 2 & 226 & 4.2 & 6.3 & 38 & 44 & 0.25 & 0.21 & 1.0 & 21 & 0.17 \\ 
			Diseaseome1 & 1 & 2 & 131 & 5.9 & 10.0 & 55 & 51 & 0.65 & 0.74 & 1.0 & 9 & 0.46 \\ 
			Diseaseome2 & 1 & 2 & 125 & 3.4 & 7.5 & 23 & 32 & 0.58 & 0.66 & 1.4 & 8 & 0.36 \\ 
			DrosophilaG & 2 & 3 & 449 & 5.9 & 4.8 & 71 & 63 & 0.31 & 0.31 & 2.0 & 34 & 0.22 \\ 
			DrosophilaG & 1 & 3 & 202 & 2.5 & 3.4 & 17 & 33 & 0.08 & 0.26 & 0.4 & 4 & 0.00 \\ 
			DrosophilaG & 1 & 2 & 299 & 2.9 & 4.1 & 25 & 49 & 0.08 & 0.31 & 0.4 & 4 & 0.11 \\ 
			DrosophilaG & 1 & 4 & 1024 & 3.4 & 7.3 & 27 & 120 & 0.05 & 0.46 & 0.5 & 9 & 0.19 \\
			HsapiensG & 4 & 5 & 422 & 3.1 & 5.7 & 48 & 27 & 0.33 & 0.38 & 1.1 & 13 & 0.26 \\ 
			HsapiensG & 2 & 4 & 913 & 18.2 & 3.0 & 813 & 105 & 0.44 & 0.28 & 1.9 & 102 & 0.28 \\ 
			HsapiensG & 1 & 4 & 881 & 9.9 & 3.1 & 186 & 91 & 0.23 & 0.31 & 1.6 & 58 & 0.22 \\ 
			HsapiensG & 2 & 6 & 363 & 9.4 & 3.0 & 103 & 73 & 0.30 & 0.22 & 0.5 & 16 & 0.31 \\ 
			HsapiensG & 2 & 3 & 179 & 7.6 & 2.4 & 45 & 25 & 0.35 & 0.03 & 0.6 & 8 & 0.00 \\ 
			HsapiensG & 1 & 6 & 293 & 7.4 & 2.9 & 62 & 65 & 0.25 & 0.25 & 0.7 & 17 & 0.14 \\ 
			HsapiensG & 2 & 5 & 4944 & 16.6 & 7.0 & 4333 & 2041 & 0.46 & 0.21 & 0.9 & 31 & 0.29 \\ 
			HsapiensG & 1 & 5 & 3886 & 9.5 & 6.9 & 789 & 1543 & 0.15 & 0.23 & 0.7 & 23 & 0.22 \\ 
			HsapiensG & 1 & 2 & 9312 & 9.1 & 14.5 & 1509 & 6715 & 0.11 & 0.40 & 2.6 & 601 & 0.15 \\ 
			HsapiensG & 1 & 3 & 169 & 5.6 & 2.3 & 50 & 28 & 0.29 & 0.04 & 0.5 & 6 & 0.00 \\ 
			Internet & 1 & 2 & 4710 & 10.2 & 5.4 & 1428 & 861 & 0.64 & 0.55 & 3.6 & 716 & 0.48 
		\end{tabular}
	\end{ruledtabular}
\end{table*}

\begin{table*}[t]
	\centering
	\caption{\label{tab:s2} Properties of the pairwise mutually connected components (MCC) for several real multiplexes. Column 2 and 3 provide the pair of layers used to construct the pairwise MCC. Column 4 provides the size of the MCC ($N_{MCC}$), which is equal to the size of the mutual graph ($\widetilde N$). Columns 5,6,11 represent the average degrees ($\langle k\rangle$), columns 7,8,11 provide the maximal degree in the network ($k_c)$ and finally, columns 9,10,12 give the average local clustering coefficients ($\langle c\rangle$).}  
	\scriptsize
	\begin{ruledtabular}
		\begin{tabular}{lll|ccccccc|ccc}
			\hline
			Multiplex &  $l$ &  $m$ & $N_{MCC}=\widetilde{N}$ &  $\langle k^{(l)}\rangle$ &  $\langle k^{(m)}\rangle$ & $k_{c}^{(l)}$  & $k_{c}^{(m)}$  & $\langle c^{(l)}\rangle$  & $\langle c^{(m)}\rangle$  &  $\langle \widetilde k\rangle$ &  $\widetilde{k}_{c}$ &  $\langle \widetilde c\rangle$ \\ \hline
			Malaria & 1 & 2 & 110 & 7.7 & 12.9 & 22 & 31 & 0.63 & 0.79 & 2.6 & 8 & 0.67 \\ 
			Malaria & 1 & 6 & 290 & 17.7 & 22.0 & 50 & 60 & 0.57 & 0.54 & 3.1 & 14 & 0.43 \\ 
			Malaria & 8 & 9 & 270 & 29.1 & 52.0 & 74 & 114 & 0.65 & 0.59 & 11.7 & 51 & 0.49 \\ 
			Malaria & 7 & 8 & 273 & 73.3 & 28.8 & 137 & 74 & 0.67 & 0.64 & 12.5 & 44 & 0.57 \\ 
			Malaria & 6 & 8 & 258 & 23.1 & 28.8 & 60 & 72 & 0.53 & 0.64 & 4.7 & 29 & 0.44 \\ 
			Malaria & 7 & 9 & 297 & 78.3 & 50.9 & 152 & 119 & 0.67 & 0.59 & 25.3 & 81 & 0.59 \\ 
			Malaria & 1 & 9 & 297 & 17.9 & 50.9 & 54 & 119 & 0.57 & 0.59 & 5.1 & 26 & 0.44 \\ 
			Malaria & 1 & 8 & 273 & 17.8 & 28.8 & 53 & 74 & 0.56 & 0.64 & 3.3 & 18 & 0.49 \\ 
			Malaria & 1 & 4 & 177 & 17.1 & 10.0 & 34 & 25 & 0.56 & 0.70 & 1.7 & 7 & 0.56 \\ 
			Malaria & 6 & 7 & 290 & 22.2 & 73.4 & 61 & 146 & 0.54 & 0.67 & 8.4 & 33 & 0.46 \\ 
			Malaria & 2 & 7 & 112 & 13.3 & 26.6 & 32 & 54 & 0.79 & 0.67 & 4.7 & 13 & 0.63 \\ 
			Malaria & 4 & 5 & 176 & 9.8 & 16.3 & 24 & 37 & 0.70 & 0.38 & 1.3 & 6 & 0.46 \\ 
			Malaria & 1 & 5 & 294 & 18.0 & 18.2 & 53 & 48 & 0.57 & 0.42 & 2.2 & 10 & 0.44 \\ 
			Malaria & 5 & 6 & 281 & 17.1 & 22.7 & 44 & 61 & 0.43 & 0.54 & 2.5 & 12 & 0.36 \\ 
			Malaria & 6 & 9 & 281 & 22.6 & 49.0 & 61 & 113 & 0.53 & 0.59 & 7.3 & 32 & 0.42 \\ 
			Malaria & 1 & 7 & 306 & 18.1 & 76.4 & 54 & 152 & 0.58 & 0.67 & 7.3 & 24 & 0.51 \\ 
			Malaria & 2 & 9 & 105 & 12.8 & 15.0 & 30 & 31 & 0.79 & 0.61 & 2.1 & 6 & 0.71 \\ 
			Malaria & 4 & 7 & 183 & 10.2 & 63.9 & 26 & 100 & 0.71 & 0.67 & 4.2 & 15 & 0.52 \\ 
			Malaria & 5 & 7 & 297 & 18.0 & 75.9 & 48 & 151 & 0.42 & 0.67 & 6.1 & 31 & 0.34 \\ 
			Malaria & 4 & 9 & 183 & 10.2 & 44.4 & 26 & 87 & 0.71 & 0.60 & 2.8 & 16 & 0.53 \\ 
			Malaria & 5 & 9 & 288 & 17.8 & 51.5 & 47 & 118 & 0.42 & 0.59 & 4.2 & 18 & 0.33 \\ 
			Malaria & 5 & 8 & 261 & 17.7 & 29.6 & 45 & 74 & 0.42 & 0.64 & 2.7 & 12 & 0.45 \\ 
			Malaria & 4 & 8 & 175 & 9.8 & 30.8 & 24 & 63 & 0.71 & 0.67 & 1.9 & 9 & 0.57 \\ 
			Malaria & 4 & 6 & 170 & 10.0 & 27.0 & 25 & 56 & 0.70 & 0.53 & 2.2 & 11 & 0.42 \\ 
			MoscowAthletics & 1 & 3 & 3424 & 5.5 & 2.7 & 583 & 245 & 0.34 & 0.07 & 1.1 & 156 & 0.06 \\ 
			MoscowAthletics & 1 & 2 & 24763 & 3.9 & 4.2 & 2679 & 3698 & 0.30 & 0.46 & 0.7 & 683 & 0.26 \\ 
			MoscowAthletics & 2 & 3 & 8210 & 6.0 & 2.5 & 1131 & 435 & 0.36 & 0.04 & 2.5 & 435 & 0.04 \\ 
			MusG & 1 & 3 & 1059 & 4.6 & 2.9 & 171 & 29 & 0.21 & 0.10 & 1.2 & 18 & 0.08 \\ 
			NYClimateMarch & 2 & 3 & 4915 & 7.4 & 2.4 & 1109 & 344 & 0.29 & 0.02 & 2.4 & 344 & 0.02 \\ 
			NYClimateMarch & 1 & 3 & 2970 & 10.6 & 2.6 & 840 & 229 & 0.23 & 0.04 & 0.8 & 116 & 0.01 \\ 
			NYClimateMarch & 1 & 2 & 37980 & 7.2 & 5.3 & 5377 & 6417 & 0.20 & 0.41 & 0.7 & 1872 & 0.26 \\ 
			Physicians & 2 & 3 & 106 & 4.3 & 3.4 & 14 & 9 & 0.23 & 0.16 & 1.4 & 7 & 0.09 \\ 
			Physicians & 1 & 2 & 104 & 4.3 & 4.3 & 22 & 14 & 0.24 & 0.24 & 2.4 & 13 & 0.24 \\ 
			RattusG & 1 & 2 & 158 & 3.0 & 2.3 & 82 & 10 & 0.34 & 0.06 & 1.1 & 6 & 0.06 \\ 
			SacchereG & 6 & 7 & 512 & 2.9 & 14.4 & 45 & 76 & 0.22 & 0.23 & 0.4 & 7 & 0.17 \\ 
			SacchereG & 1 & 6 & 677 & 16.4 & 2.9 & 501 & 51 & 0.47 & 0.24 & 1.4 & 17 & 0.29 \\ 
			SacchereG & 2 & 7 & 4335 & 14.6 & 40.5 & 624 & 454 & 0.11 & 0.12 & 1.0 & 46 & 0.15 \\ 
			SacchereG & 3 & 6 & 503 & 6.8 & 2.9 & 68 & 35 & 0.29 & 0.27 & 1.2 & 10 & 0.20 \\ 
			SacchereG & 3 & 5 & 729 & 7.5 & 3.5 & 106 & 79 & 0.31 & 0.43 & 1.4 & 18 & 0.29 \\ 
			SacchereG & 4 & 6 & 525 & 10.9 & 2.9 & 108 & 37 & 0.28 & 0.24 & 0.6 & 7 & 0.23 \\ 
			SacchereG & 2 & 6 & 468 & 7.4 & 2.9 & 44 & 42 & 0.19 & 0.22 & 0.7 & 7 & 0.09 \\ 
			SacchereG & 5 & 7 & 413 & 3.2 & 17.3 & 62 & 94 & 0.40 & 0.27 & 0.4 & 5 & 0.10 \\ 
			SacchereG & 2 & 5 & 609 & 7.2 & 3.4 & 41 & 65 & 0.17 & 0.42 & 0.7 & 12 & 0.24 \\ 
			SacchereG & 1 & 2 & 4531 & 20.6 & 14.1 & 2738 & 613 & 0.43 & 0.11 & 0.9 & 25 & 0.30 \\ 
			SacchereG & 1 & 7 & 4720 & 18.4 & 37.4 & 2804 & 457 & 0.42 & 0.12 & 0.8 & 72 & 0.18 \\ 
			SacchereG & 1 & 5 & 800 & 25.3 & 3.5 & 541 & 97 & 0.51 & 0.43 & 2.5 & 35 & 0.49 \\ 
			SacchereG & 4 & 5 & 660 & 11.8 & 3.5 & 168 & 70 & 0.32 & 0.43 & 0.7 & 9 & 0.25 \\ 
			SacchereG & 2 & 4 & 4019 & 14.9 & 15.9 & 524 & 966 & 0.12 & 0.28 & 1.1 & 46 & 0.15 \\ 
			SacchereG & 2 & 3 & 3954 & 14.1 & 11.2 & 510 & 392 & 0.12 & 0.16 & 0.8 & 27 & 0.21 \\ 
			SacchereG & 4 & 7 & 4111 & 15.2 & 39.9 & 1006 & 445 & 0.29 & 0.13 & 3.4 & 142 & 0.23 \\ 
			SacchereG & 1 & 4 & 4494 & 22.3 & 14.8 & 2701 & 1059 & 0.42 & 0.28 & 0.9 & 55 & 0.27 \\ 
			SacchereG & 1 & 3 & 4571 & 21.6 & 11.2 & 2769 & 435 & 0.42 & 0.16 & 1.9 & 190 & 0.31 \\ 
			SacchereG & 3 & 7 & 4065 & 10.9 & 36.6 & 382 & 435 & 0.16 & 0.13 & 0.5 & 21 & 0.15 \\ 
			SacchereG& 3 & 4 & 3903 & 11.1 & 15.6 & 390 & 933 & 0.16 & 0.28 & 0.7 & 53 & 0.19 \\ 
			SacchpombG & 1 & 5 & 289 & 3.0 & 6.2 & 17 & 37 & 0.20 & 0.32 & 0.9 & 7 & 0.29 \\ 
			SacchpombG & 5 & 6 & 526 & 6.5 & 18.1 & 43 & 119 & 0.24 & 0.23 & 1.8 & 19 & 0.17 \\ 
			SacchpombG & 1 & 4 & 424 & 3.2 & 6.3 & 23 & 31 & 0.15 & 0.19 & 0.9 & 6 & 0.20 \\ 
			SacchpombG& 4 & 5 & 646 & 7.6 & 6.4 & 55 & 50 & 0.15 & 0.27 & 0.8 & 14 & 0.18 \\ 
			SacchpombG& 1 & 6 & 332 & 3.0 & 9.8 & 20 & 58 & 0.16 & 0.19 & 0.3 & 6 & 0.00 \\ 
			SacchpombG& 1 & 3 & 510 & 3.2 & 4.5 & 25 & 101 & 0.20 & 0.35 & 1.5 & 14 & 0.21 \\ 
			SacchpombG & 3 & 5 & 426 & 3.9 & 5.2 & 48 & 36 & 0.32 & 0.25 & 0.8 & 9 & 0.15 \\ 
			SacchpombG & 3 & 4 & 1112 & 4.9 & 8.5 & 428 & 132 & 0.40 & 0.12 & 0.5 & 8 & 0.30 \\ 
			SacchpombG & 4 & 6 & 2292 & 12.5 & 25.8 & 312 & 347 & 0.07 & 0.13 & 0.2 & 10 & 0.06 \\ 
			SacchpombG & 3 & 6 & 956 & 4.9 & 19.5 & 414 & 167 & 0.44 & 0.16 & 0.3 & 11 & 0.04 
		\end{tabular}
	\end{ruledtabular}
\end{table*}

\begin{table*}[t]
	\caption{\label{tab:s3} For all 3 pairs of graphs from the set $\{G_l,G_m,\widetilde{G}\}$ we compare the degree sequences ($\boldsymbol{k}$) as well as the local clustering coefficient sequences ($\boldsymbol{c}$) using the Spearman's ranking correlation coefficient $r_s$. Results are not shown when either of the two sequences being compared are constant.}
	\centering
	\begin{ruledtabular}
		\scriptsize
		\begin{tabular}{lcc|cccccc}
			Multiplex & $l$ & $m$ &  $r_s(\boldsymbol{k^{(l)}},\boldsymbol{k^{(m)}})$ &  $r_s(\boldsymbol{k^{(l)}},\boldsymbol{\widetilde k})$ &  $r_s(\boldsymbol{k^{(m)}},\boldsymbol{\widetilde k})$ &  $r_s(\boldsymbol{c^{(l)}},\boldsymbol{c^{(m)}})$ &  $r_s(\boldsymbol{c^{(l)}},\boldsymbol{\widetilde c})$ &  $r_s(\boldsymbol{c^{(m)}},\boldsymbol{\widetilde c})$ \\
			\hline
			ArabidopsisG &      1 &      2 &     0.20 &     0.63 &     0.11 &     0.19 &     0.50 &     0.13 \\
			arXiv &      2 &      3 &     0.29 &     0.77 &     0.18 &     0.05 &     0.62 &     0.10 \\
			arXiv &      2 &      5 &     0.21 &     0.85 &     0.17 &     0.13 &     0.80 &     0.11 \\
			arXiv &      3 &      5 &     0.16 &     0.78 &     0.04 &     0.09 &     0.48 &     0.19 \\
			arXiv &      2 &      6 &     0.29 &     0.82 &     0.20 &    -0.00 &     0.67 &     0.01 \\
			arXiv &      3 &      6 &     0.24 &     0.93 &     0.20 &     0.04 &     0.80 &     0.01 \\
			arXiv &      5 &      6 &     0.31 &     0.79 &     0.33 &     0.11 &     0.62 &    -0.01 \\
			arXiv &      2 &      7 &     0.35 &     0.86 &     0.38 &     0.13 &     0.73 &     0.15 \\
			arXiv &      2 &     12 &     0.27 &     0.95 &     0.23 &     0.14 &     0.90 &     0.11 \\
			arXiv &      3 &     12 &     0.23 &     0.70 &     0.13 &     0.09 &     0.49 &     0.11 \\
			arXiv &      5 &     12 &     0.28 &     0.89 &     0.25 &     0.15 &     0.72 &     0.16 \\
			arXiv &      6 &     12 &     0.31 &     0.76 &     0.23 &     0.03 &     0.59 &    -0.03 \\
			CelegansC &      1 &      2 &     0.18 &     0.36 &     0.12 &     0.22 &     0.22 &     0.04 \\
			CelegansC &      1 &      3 &     0.18 &     0.49 &     0.14 &     0.15 &     0.47 &     0.04 \\
			CelegansC &      2 &      3 &     0.27 &     0.89 &     0.25 &     0.09 &     0.75 &     0.05 \\
			Diseaseome1 &      1 &      2 &     0.35 &     0.67 &     0.11 &    -0.01 &     0.01 &    -0.01 \\
			Diseaseome2 &      1 &      2 &     0.44 &     0.63 &     0.24 &     0.18 &     0.17 &    -0.01 \\
			DrosophilaG &      1 &      2 &     0.21 &     0.38 &     0.14 &    -0.02 &     0.23 &     0.02 \\
			DrosophilaG &      1 &      3 &     0.11 &     0.40 &     0.07 &    -0.03 &      -- &      -- \\
			DrosophilaG &      2 &      3 &     0.24 &     0.72 &     0.15 &     0.09 &     0.27 &     0.05 \\
			DrosophilaG &      1 &      4 &     0.23 &     0.32 &     0.05 &     0.04 &     0.47 &    -0.01 \\
			HsapiensG &      1 &      2 &     0.39 &     0.65 &     0.24 &     0.08 &     0.48 &     0.03 \\
			HsapiensG &      1 &      3 &    -0.14 &     0.56 &    -0.06 &    -0.11 &      -- &      -- \\
			HsapiensG &      2 &      3 &     0.39 &     0.57 &     0.32 &     0.01 &      -- &      -- \\
			HsapiensG &      1 &      4 &     0.27 &     0.49 &     0.13 &     0.02 &     0.21 &     0.02 \\
			HsapiensG &      2 &      4 &     0.29 &     0.51 &     0.14 &    -0.11 &    -0.07 &     0.03 \\
			HsapiensG &      1 &      5 &     0.26 &     0.51 &     0.10 &     0.12 &     0.24 &     0.03 \\
			HsapiensG &      2 &      5 &     0.50 &     0.56 &     0.30 &    -0.01 &    -0.02 &     0.10 \\
			HsapiensG &      4 &      5 &     0.14 &     0.59 &     0.20 &     0.15 &     0.46 &     0.11 \\
			HsapiensG &      1 &      6 &     0.25 &     0.52 &     0.21 &     0.11 &     0.11 &     0.08 \\
			HsapiensG &      2 &      6 &     0.27 &     0.46 &     0.13 &     0.10 &     0.10 &     0.18 \\
			Internet &      1 &      2 &     0.27 &     0.67 &     0.20 &     0.09 &     0.24 &     0.10 \\
		\end{tabular}
	\end{ruledtabular}
\end{table*}

\begin{table*}[t]
	\caption{\label{tab:s4} For all 3 pairs of graphs from the set $\{G_l,G_m,\widetilde{G}\}$ we compare the degree sequences ($\boldsymbol{k}$) as well as the local clustering coefficient sequences ($\boldsymbol{c}$) using the Spearman's ranking correlation coefficient $r_s$.}
	\centering
	\begin{ruledtabular}
		\scriptsize
		\begin{tabular}{lcc|cccccc}
			Multiplex & $l$ & $m$ &  $r_s(\boldsymbol{k^{(l)}},\boldsymbol{k^{(m)}})$ &  $r_s(\boldsymbol{k^{(l)}},\boldsymbol{\widetilde k})$ &  $r_s(\boldsymbol{k^{(m)}},\boldsymbol{\widetilde k})$ &  $r_s(\boldsymbol{c^{(l)}},\boldsymbol{c^{(m)}})$ &  $r_s(\boldsymbol{c^{(l)}},\boldsymbol{\widetilde c})$ &  $r_s(\boldsymbol{c^{(m)}},\boldsymbol{\widetilde c})$ \\
			\hline
			Internet &      1 &      2 &     0.27 &     0.67 &     0.20 &     0.09 &     0.24 &     0.10 \\
			Malaria &      1 &      2 &     0.16 &     0.38 &     0.12 &     0.18 &     0.38 &     0.27 \\
			Malaria &      1 &      4 &     0.19 &     0.36 &     0.20 &     0.04 &     0.13 &     0.08 \\
			Malaria &      1 &      5 &     0.02 &     0.49 &    -0.02 &    -0.03 &     0.23 &     0.04 \\
			Malaria &      4 &      5 &     0.22 &     0.39 &    -0.05 &    -0.10 &    -0.19 &     0.02 \\
			Malaria &      1 &      6 &     0.29 &     0.54 &     0.25 &     0.03 &     0.04 &    -0.03 \\
			Malaria &      4 &      6 &     0.31 &     0.56 &     0.19 &    -0.08 &     0.01 &     0.20 \\
			Malaria &      5 &      6 &    -0.01 &     0.49 &     0.01 &     0.10 &     0.11 &    -0.01 \\
			Malaria &      1 &      7 &     0.30 &     0.72 &     0.30 &    -0.08 &     0.32 &     0.03 \\
			Malaria &      2 &      7 &    -0.05 &     0.54 &     0.25 &    -0.02 &     0.36 &     0.01 \\
			Malaria &      4 &      7 &    -0.10 &     0.75 &     0.01 &    -0.01 &     0.19 &     0.02 \\
			Malaria &      5 &      7 &     0.10 &     0.74 &     0.17 &     0.03 &     0.26 &     0.03 \\
			Malaria &      6 &      7 &     0.40 &     0.85 &     0.35 &    -0.08 &     0.50 &     0.04 \\
			Malaria &      1 &      8 &     0.33 &     0.56 &     0.13 &     0.04 &     0.15 &     0.10 \\
			Malaria &      4 &      8 &     0.04 &     0.44 &    -0.01 &     0.03 &     0.00 &    -0.03 \\
			Malaria &      5 &      8 &    -0.05 &     0.43 &    -0.02 &    -0.02 &     0.18 &     0.01 \\
			Malaria &      6 &      8 &     0.31 &     0.74 &     0.24 &     0.04 &     0.21 &     0.12 \\
			Malaria &      7 &      8 &     0.22 &     0.62 &     0.08 &     0.01 &     0.20 &    -0.02 \\
			Malaria &      1 &      9 &     0.10 &     0.59 &    -0.01 &     0.07 &     0.16 &    -0.08 \\
			Malaria &      2 &      9 &     0.25 &     0.36 &     0.16 &     0.02 &     0.25 &    -0.09 \\
			Malaria &      4 &      9 &    -0.12 &     0.45 &     0.00 &    -0.04 &     0.05 &    -0.00 \\
			Malaria &      5 &      9 &     0.08 &     0.62 &    -0.11 &     0.08 &     0.22 &    -0.11 \\
			Malaria &      6 &      9 &     0.01 &     0.80 &     0.06 &    -0.03 &     0.35 &     0.01 \\
			Malaria &      7 &      9 &    -0.14 &     0.67 &    -0.12 &    -0.06 &     0.54 &     0.00 \\
			Malaria &      8 &      9 &     0.16 &     0.83 &     0.14 &     0.01 &     0.39 &     0.06 \\
			MoscowAthletics &      1 &      2 &     0.10 &     0.45 &     0.04 &     0.10 &     0.29 &     0.03 \\
			MoscowAthletics &      1 &      3 &     0.08 &     0.38 &     0.04 &     0.08 &     0.10 &     0.03 \\
			MoscowAthletics &      2 &      3 &     0.05 &     0.70 &     0.02 &     0.05 &     0.18 &     0.01 \\
			MusG &      1 &      3 &     0.21 &     0.51 &     0.13 &     0.06 &     0.23 &     0.01 \\
			NYClimateMarch2014 &      1 &      2 &     0.25 &     0.48 &     0.14 &     0.19 &     0.23 &     0.09 \\
			NYClimateMarch2014 &      1 &      3 &     0.14 &     0.36 &     0.05 &     0.08 &     0.03 &     0.04 \\
			NYClimateMarch2014 &      2 &      3 &     0.15 &     0.67 &     0.07 &     0.08 &     0.11 &     0.05 \\
			Physicians &      1 &      2 &     0.19 &     0.71 &     0.12 &    -0.17 &     0.48 &     0.12 \\
			Physicians &      2 &      3 &     0.13 &     0.53 &     0.22 &     0.18 &     0.15 &    -0.06 \\
			RattusG &      1 &      2 &     0.39 &     0.63 &     0.28 &     0.13 &     0.14 &     0.30 \\
			SacchereG &      1 &      2 &     0.50 &     0.52 &     0.27 &    -0.03 &    -0.06 &     0.10 \\
			SacchereG &      1 &      3 &     0.46 &     0.61 &     0.27 &     0.02 &    -0.06 &     0.08 \\
			SacchereG &      2 &      3 &     0.40 &     0.47 &     0.19 &     0.12 &     0.23 &     0.06 \\
			SacchereG &      1 &      4 &     0.49 &     0.52 &     0.26 &     0.04 &    -0.07 &     0.01 \\
			SacchereG &      2 &      4 &     0.41 &     0.50 &     0.21 &     0.09 &     0.21 &     0.01 \\
			SacchereG &      3 &      4 &     0.39 &     0.44 &     0.20 &     0.09 &     0.15 &    -0.01 \\
			SacchereG &      1 &      5 &     0.15 &     0.38 &     0.06 &     0.03 &     0.11 &     0.05 \\
			SacchereG &      2 &      5 &     0.20 &     0.39 &     0.08 &     0.08 &     0.24 &    -0.06 \\
			SacchereG &      3 &      5 &     0.15 &     0.48 &    -0.01 &    -0.00 &     0.25 &    -0.02 \\
			SacchereG &      4 &      5 &     0.15 &     0.39 &     0.02 &    -0.06 &     0.08 &    -0.05 \\
			SacchereG &      1 &      6 &     0.24 &     0.46 &     0.12 &    -0.00 &     0.03 &    -0.00 \\
			SacchereG &      2 &      6 &     0.20 &     0.42 &     0.07 &     0.10 &     0.10 &     0.02 \\
			SacchereG &      3 &      6 &     0.16 &     0.44 &     0.08 &    -0.01 &     0.29 &     0.05 \\
			SacchereG &      4 &      6 &     0.17 &     0.43 &     0.03 &     0.11 &     0.12 &     0.04 \\
			SacchereG &      1 &      7 &     0.56 &     0.55 &     0.29 &    -0.01 &    -0.03 &     0.07 \\
			SacchereG &      2 &      7 &     0.47 &     0.59 &     0.28 &     0.11 &     0.22 &     0.08 \\
			SacchereG &      3 &      7 &     0.44 &     0.44 &     0.22 &     0.14 &     0.12 &     0.03 \\
			SacchereG &      4 &      7 &     0.49 &     0.69 &     0.30 &     0.08 &     0.18 &     0.09 \\
			SacchereG &      5 &      7 &     0.04 &     0.43 &    -0.01 &    -0.03 &     0.13 &    -0.06 \\
			SacchereG &      6 &      7 &     0.13 &     0.33 &     0.08 &     0.09 &     0.23 &     0.04 \\
			SacchpombG &      1 &      3 &     0.30 &     0.67 &     0.21 &     0.09 &     0.58 &    -0.02 \\
			SacchpombG &      1 &      4 &     0.36 &     0.51 &     0.24 &     0.14 &     0.44 &     0.04 \\
			SacchpombG &      3 &      4 &     0.20 &     0.41 &     0.09 &     0.05 &     0.21 &     0.05 \\
			SacchpombG &      1 &      5 &     0.29 &     0.47 &     0.16 &     0.10 &     0.48 &     0.09 \\
			SacchpombG &      3 &      5 &     0.24 &     0.49 &     0.18 &     0.10 &     0.24 &     0.05 \\
			SacchpombG &      4 &      5 &     0.36 &     0.37 &     0.30 &     0.14 &     0.25 &     0.14 \\
			SacchpombG &      1 &      6 &     0.30 &     0.35 &     0.17 &     0.10 &      -- &      -- \\
			SacchpombG &      3 &      6 &     0.28 &     0.35 &     0.10 &    -0.02 &     0.01 &     0.01 \\
			SacchpombG &      4 &      6 &     0.33 &     0.23 &     0.19 &     0.14 &     0.09 &     0.01 \\
			SacchpombG &      5 &      6 &     0.38 &     0.65 &     0.25 &     0.11 &     0.31 &     0.04 \\
		\end{tabular}
	\end{ruledtabular}
\end{table*}

\begin{table*}[t]
	\caption{\label{tab:s5} The geometric coupling coefficients $\beta$ and mean distance between connected nodes $\langle d\rangle_E$ for layers $l$ and $m$ as well as for the giant connected component of the mutual graph. The first two rows represent the size of the giant connected component of the MG $\widetilde{N}_{GCC}$ as well its ratio with the full size of the MG $\widetilde{N}$.}
	\centering
	\begin{ruledtabular}
		\scriptsize
		\begin{tabular}{lcc|cccccccc}
			\hline
			Multiplex &  $l$ & $ m$ &$\widetilde{N}_{GCC}$ & $\widetilde{N}_{GCC}/\widetilde{N}$ &$\beta_l$ &  $\beta_m$ &  $\widetilde\beta$ &$\langle d^{(1)}\rangle_E$&$\langle d^{(2)}\rangle_E$&$\langle \widetilde d\rangle_E$\\ \hline
ArabidopsisG &      1 &      2 &   144 &   0.33 &  1.37 &  1.57 &  1.43 & 0.24 & 0.20 & 0.11 \\
arXiv &      2 &      3 &   445 &   0.79 &  3.73 &  8.00 & 29.36 & 0.32 & 0.18 & 0.14 \\
arXiv &      2 &      5 &   486 &   0.96 &  6.54 & 33.99 & 26.82 & 0.27 & 0.14 & 0.14 \\
arXiv &      2 &      6 &   753 &   0.82 &  3.50 &  5.75 & 16.28 & 0.34 & 0.19 & 0.16 \\
arXiv &      3 &      6 &   736 &   0.93 &  6.09 &  4.98 & 16.92 & 0.16 & 0.21 & 0.12 \\
arXiv &      5 &      6 &    84 &   0.46 & 13.28 &  7.10 & 31.11 & 0.15 & 0.25 & 0.30 \\
arXiv &      2 &      7 &   182 &   0.87 & 34.50 & 26.25 & 27.91 & 0.28 & 0.11 & 0.11 \\
arXiv &      2 &     12 &  2164 &   0.96 &  9.94 & 28.35 & 29.21 & 0.24 & 0.23 & 0.18 \\
arXiv &      3 &     12 &   124 &   0.42 & 11.18 &  4.68 & 27.29 & 0.17 & 0.17 & 0.17 \\
arXiv &      5 &     12 &   227 &   0.73 & 30.31 &  6.24 & 25.89 & 0.10 & 0.19 & 0.10 \\
arXiv &      6 &     12 &   328 &   0.63 &  5.59 &  4.40 & 29.76 & 0.18 & 0.21 & 0.12 \\
Celegans\_G &      1 &      3 &    91 &   0.37 &  1.34 &  1.28 &  1.37 & 0.32 & 0.67 & 0.30 \\
Celegans\_G &      2 &      3 &   242 &   0.94 &  1.18 &  1.37 &  1.18 & 0.50 & 0.69 & 0.45 \\
DrosophilaG &      2 &      3 &   261 &   0.58 &  1.38 &  1.47 &  1.31 & 0.52 & 0.46 & 0.23 \\
HsapiensG &      1 &      2 &  4958 &   0.53 &  0.92 &  1.02 &  1.12 & 0.87 & 0.95 & 0.56 \\
HsapiensG &      1 &      4 &   458 &   0.52 &  1.15 &  1.68 &  1.39 & 0.84 & 0.14 & 0.09 \\
HsapiensG &      2 &      4 &   601 &   0.66 &  1.30 &  1.56 &  1.67 & 0.96 & 0.13 & 0.08 \\
HsapiensG &      1 &      5 &   653 &   0.17 &  0.96 &  0.94 &  1.46 & 0.88 & 0.41 & 0.11 \\
HsapiensG &      2 &      5 &  1057 &   0.21 &  1.08 &  0.80 &  1.53 & 1.01 & 0.39 & 0.16 \\
Internet &      1 &      2 &  4260 &   0.90 &  1.58 &  1.41 &  1.45 & 0.75 & 0.64 & 0.56 \\
Malaria &      1 &      5 &   147 &   0.50 &  2.89 &  1.85 &  1.99 & 0.34 & 0.83 & 0.16 \\
Malaria &      1 &      6 &   178 &   0.61 &  2.90 &  2.20 &  2.12 & 0.35 & 0.52 & 0.28 \\
Malaria &      4 &      6 &    70 &   0.41 &  5.26 &  2.50 &  2.07 & 0.25 & 0.63 & 0.27 \\
Malaria &      5 &      6 &   149 &   0.53 &  1.88 &  2.18 &  1.66 & 0.84 & 0.53 & 0.39 \\
Malaria &      1 &      7 &   273 &   0.89 &  2.92 &  3.95 &  2.27 & 0.34 & 0.81 & 0.21 \\
Malaria &      2 &      7 &    95 &   0.85 & 34.59 &  4.06 &  3.53 & 0.49 & 0.70 & 0.25 \\
Malaria &      4 &      7 &   150 &   0.82 &  5.26 &  4.32 &  2.67 & 0.27 & 0.78 & 0.19 \\
Malaria &      5 &      7 &   239 &   0.80 &  1.84 &  4.04 &  1.59 & 0.84 & 0.78 & 0.54 \\
Malaria &      6 &      7 &   197 &   0.68 &  2.21 &  4.02 &  1.82 & 0.51 & 0.83 & 0.57 \\
Malaria &      1 &      8 &    94 &   0.34 &  2.68 &  2.58 &  2.00 & 0.36 & 0.64 & 0.38 \\
Malaria &      4 &      8 &    54 &   0.31 &  5.73 &  3.36 &  3.54 & 0.25 & 0.50 & 0.22 \\
Malaria &      5 &      8 &   118 &   0.45 &  1.82 &  2.59 &  1.77 & 0.95 & 0.57 & 0.42 \\
Malaria &      6 &      8 &   129 &   0.50 &  2.12 &  2.60 &  1.85 & 0.56 & 0.46 & 0.54 \\
Malaria &      7 &      8 &   232 &   0.85 &  4.18 &  2.57 &  2.20 & 0.73 & 0.62 & 0.52 \\
Malaria &      1 &      9 &   216 &   0.73 &  2.80 &  2.43 &  1.91 & 0.33 & 0.66 & 0.30 \\
Malaria &      4 &      9 &    82 &   0.45 &  5.54 &  2.49 &  2.49 & 0.27 & 0.71 & 0.25 \\
Malaria &      5 &      9 &   210 &   0.73 &  1.82 &  2.44 &  1.62 & 0.89 & 0.67 & 0.44 \\
Malaria &      6 &      9 &   184 &   0.65 &  2.17 &  2.39 &  1.55 & 0.53 & 0.69 & 0.68 \\
Malaria &      7 &      9 &   268 &   0.90 &  4.10 &  2.43 &  2.24 & 0.85 & 0.67 & 0.51 \\
Malaria &      8 &      9 &   185 &   0.69 &  2.72 &  2.39 &  1.88 & 0.53 & 0.69 & 0.59 \\
MoscowAthletics &      1 &      2 &  5254 &   0.21 &  1.12 &  1.40 &  1.15 & 0.25 & 0.37 & 0.23 \\
MoscowAthletics &      1 &      3 &  1254 &   0.37 &  1.31 &  0.60 &  0.00 & 0.48 & 0.18 & 0.11 \\
MoscowAthletics &      2 &      3 &  8210 &   1.00 &  1.24 &  0.00 &  0.00 & 0.47 & 0.15 & 0.15 \\
MusG &      1 &      3 &   352 &   0.33 &  1.23 &  1.12 &  0.92 & 0.44 & 0.19 & 0.08 \\
NYClimateMarch2014 &      1 &      2 &  8153 &   0.21 &  0.69 &  1.17 &  1.00 & 0.82 & 0.66 & 0.44 \\
NYClimateMarch2014 &      1 &      3 &   717 &   0.24 &  0.95 &  0.00 &  0.00 & 0.76 & 0.26 & 0.11 \\
NYClimateMarch2014 &      2 &      3 &  4915 &   1.00 &  1.12 &  0.00 &  0.00 & 0.80 & 0.22 & 0.22 \\
Physicians &      1 &      2 &    85 &   0.82 &  1.30 &  1.39 &  1.48 & 0.59 & 0.51 & 0.29 \\
SacchereG &      1 &      2 &  1091 &   0.24 &  1.06 &  0.89 &  1.71 & 0.88 & 0.79 & 0.13 \\
SacchereG &      1 &      3 &  2158 &   0.47 &  1.08 &  1.03 &  1.66 & 0.94 & 0.76 & 0.24 \\
SacchereG &      2 &      3 &   812 &   0.21 &  0.91 &  1.04 &  1.54 & 0.85 & 0.71 & 0.11 \\
SacchereG &      1 &      4 &  1086 &   0.24 &  1.06 &  0.81 &  1.53 & 0.87 & 0.88 & 0.21 \\
SacchereG &      2 &      4 &  1132 &   0.28 &  0.91 &  0.85 &  1.22 & 0.80 & 0.93 & 0.26 \\
SacchereG &      3 &      4 &   839 &   0.21 &  1.05 &  0.84 &  1.36 & 0.73 & 0.89 & 0.18 \\
SacchereG &      1 &      5 &   182 &   0.23 &  1.89 &  2.07 &  2.97 & 0.74 & 0.06 & 0.17 \\
SacchereG &      3 &      5 &    85 &   0.12 &  1.50 &  2.06 &  1.87 & 0.51 & 0.09 & 0.19 \\
SacchereG &      1 &      6 &    60 &   0.09 &  1.40 &  1.51 &  1.76 & 0.90 & 0.07 & 0.20 \\
SacchereG &      3 &      6 &    55 &   0.11 &  1.48 &  1.58 &  1.49 & 0.58 & 0.07 & 0.18 \\
SacchereG &      1 &      7 &  1026 &   0.22 &  0.99 &  0.73 &  1.33 & 0.90 & 1.03 & 0.19 \\
SacchereG &      2 &      7 &  1127 &   0.26 &  0.88 &  0.74 &  1.21 & 0.79 & 1.04 & 0.22 \\
SacchereG &      3 &      7 &   663 &   0.16 &  1.04 &  0.72 &  1.24 & 0.73 & 1.02 & 0.10 \\
Sacchere &      4 &      7 &  1584 &   0.39 &  0.83 &  0.75 &  1.05 & 0.93 & 1.02 & 0.60 \\
SacchpombG &      1 &      3 &   186 &   0.36 &  1.41 &  1.59 &  1.23 & 0.20 & 0.47 & 0.10 \\
SacchpombG &      1 &      4 &    60 &   0.14 &  1.24 &  1.17 &  1.19 & 0.23 & 0.57 & 0.21 \\
SacchpombG &      4 &      5 &   160 &   0.25 &  1.06 &  1.35 &  1.38 & 0.72 & 0.48 & 0.12 \\
SacchpombG &      4 &      6 &   107 &   0.05 &  0.00 &  0.00 &  0.00 & 1.07 & 1.14 & 0.21 \\
SacchpombG &      5 &      6 &   228 &   0.43 &  1.23 &  0.63 &  1.07 & 0.53 & 0.87 & 0.44 
		\end{tabular}
	\end{ruledtabular}
\end{table*}

\clearpage 

\section{The average mutual degree}
In this section we study the scaling behaviors of the average mutual degree, given by the following equation:
\begin{equation}
	\langle \widetilde{k} \rangle = N \iint d^L\boldsymbol{h}d^L\boldsymbol{h'}\rho(\boldsymbol{h})\rho(\boldsymbol{h'})\widetilde{p}(\boldsymbol{h},\boldsymbol{h'})\label{eq:M_generalh}, 
\end{equation} 
where $d^L \boldsymbol{h} = \prod_{l=1}^{L}dh^{(l)}$, $\rho(\boldsymbol{h}) = \rho(h^{(1)})\prod_{l=2}^L \rho\left(h^{(l)}|h^{(l-1)}\right)$ and where 
\begin{equation}
	\widetilde{p}(\boldsymbol{h},\boldsymbol{h'}) \equiv P (\boldsymbol{h}\sim \boldsymbol{h'}) =\prod_{l=1}^{L} p_l(h^{(l)},h'^{(l)})\label{eq:hsimh'}.
\end{equation}

\subsection{Perfectly correlated layers, homogeneous degree distribution}
We start with the case of $L$ sparse, homogeneous, perfectly correlated layers with general $\{\beta_l\}_{l=1}^L$ and $\{\langle k\rangle^{(l)}\}_{l=1}^L$. In this setting, $\boldsymbol{h}=\{\theta^{(l)}\}_{l=1}^L$ and  $\rho(\theta^{(l)}|\theta^{(l-1)})=\delta(\theta^{(l)}-\theta^{(l-1)})$ $\forall l$, implying that Eq.~\eqref{eq:M_generalh} can be written as
\begin{equation}
	\langle \widetilde{k}\rangle=\frac{N}{\pi}\int_0^\pi d\theta\prod_{i=1}^L\frac{1}{1+(\zeta_i\theta)^{\beta_i}}.
\end{equation}
Without loss of generality, we now assume that $\zeta_1\geq \zeta_2\geq ...\geq \zeta_L$. Note that we want to have sparse individual layers such that $\zeta_i = N/(2\pi\mu^{(i)}\langle k^{(i)}\rangle^2) \sim N^{\max(1,1/\beta_i)}$. This implies that $\beta_1\leq \beta_2\leq ...\leq \beta_L$ when $N\gg1$. This allows us to split the integral in $L+1$ intervals
\begin{equation}
	\langle \widetilde{N}\rangle = \frac{N}{\pi}\sum_{j=1}^{L+1}I_j,
\end{equation}
where 
\begin{equation}
	I_j=\int_{\zeta_{j-1}^{-1}}^{\zeta_j^{-1}} d\theta\prod_{i=1}^L\frac{1}{1+(\zeta_i\theta)^{\beta_i}}
\end{equation}
and where we have defined $\zeta_0^{-1}=0$ and $\zeta_{L+1}^{-1}=\pi$. We can now find the upper and lower bound of each $I_j$ separately. We start with the lower bound:
{
	\begin{alignat}{6}
		I_j\geq \int_{\zeta_{j-1}^{-1}}^{\zeta_j^{-1}} d\theta\min_\theta\left(\prod_{i=1}^L\frac{1}{1+(\zeta_i\theta)^{\beta_i}}\right)&=\left(\zeta_{j}^{-1}-\zeta_{j-1}^{-1}\right)\prod_{i=1}^L\frac{1}{1+(\zeta_i/\zeta_j)^{\beta_i}}\notag\\[3mm]
		&\simeq \zeta_j^{-1}\prod_{i=1}^{j-1}\left(\frac{\zeta_i}{\zeta_j}\right)^{-\beta_i}.
\end{alignat}}
Here we make use of the fact that $\zeta_{j-1}\geq \zeta_j$ so $\zeta^{-1}_{j-1}\leq \zeta^{-1}_j$. Furthermore, the fraction $\zeta_i/\zeta_j$ is large only for $i<j$, and will thus only then show up in the Taylor expansion. Otherwise the fraction can just be approximated as $1$. For the upper bound we will use the fact that $1\geq 1/(1+x)$ and $1\geq 1/x$. Now, we want to make as tight a bound as possible, so when $1\geq 1/x$ we take $1/x$ and when $1\leq1/x$ we take $1$. We thus only have a contribution to the integrand when $x\geq1$. In our case this implies $\zeta_i\theta\geq1$. This is the case for the entire integration range only when $i\leq j-1$. Thus, we can write the upper bound as 
{
	\begin{equation}
		I_j\leq \int_{\zeta_{j-1}^{-1}}^{\zeta_j^{-1}} d\theta\prod_{i=1}^{j-1}\left(\zeta_i\theta\right)^{-\beta_i}=\frac{\zeta_j^{-1}\prod_{i=1}^{j-1}\left(\frac{\zeta_i}{\zeta_j}\right)^{-\beta_i}-\zeta_{j-1}^{-1}\prod_{i=1}^{j-1}\left(\frac{\zeta_i}{\zeta_{j-1}}\right)^{-\beta_i}}{1-\sum_{i=1}^{j-1}\beta_i}.
\end{equation} }
The product in the second term in the numerator can be taken instead to $j-2$ because the $j-1$ term just evaluates to one. Then, one sees that both terms are equivalent (ignoring prefactors) if for the second term one perform $j-1\rightarrow j$. Thus, when summing over $I_j$, the second contribution will already come from the $I_{j-1}$ integral. We are left with
\begin{equation}
	\sum_j\left(\zeta_j^{-1}\prod_{i=1}^{j-1}\left(\frac{\zeta_i}{\zeta_j}\right)^{-\beta_i}\right)\leq \sum_jI_j\leq \sum_j\left(\zeta_j^{-1}\prod_{i=1}^{j-1}\left(\frac{\zeta_i}{\zeta_j}\right)^{-\beta_i}\right).
\end{equation}
We have then proven that 
\begin{equation}
	\langle \widetilde{k}\rangle \sim \frac{N}{\pi}\sum_{j=1}^{L+1}\left(\zeta_j^{-1}\prod_{i=1}^{j-1}\left(\frac{\zeta_i}{\zeta_j}\right)^{-\beta_i}\right)\sim N^{\max_{j}(\sigma_j)}.
\end{equation}
Now, the question is to find out for which $\sigma_j$ dominates. Let us ask study $\sigma_j$ and $\sigma_{j-1}$. The difference between these two exponents is given by 
\begin{equation}
	\sigma_j-\sigma_{j-1} = \left(\max\left(1,\frac{1}{\beta_j}\right)-\max\left(1,\frac{1}{\beta_{j-1}}\right)\right)\left(\sum_{i=1}^{j-1}\beta_i-1\right),
\end{equation} 
where we set $\beta_{L+1}\geq1$. Now, we know that $\zeta_{j-1}\geq \zeta_j$, which implies that $\beta_{j-1}\leq\beta_j$. This means that $\left(\max\left(1,\frac{1}{\beta_j}\right)-\max\left(1,\frac{1}{\beta_{j-1}}\right)\right)\leq0$ and so
\begin{equation}
	\sigma_j\geq\sigma_{j-1}\quad\text{if}\quad\sum_{i=1}^{j-1}\beta_i\leq1.
\end{equation}
Thus, the largest exponent is given by the $j$ that satisfies 
\begin{equation}
	1-\beta_j\leq\sum_{i=1}^{j-1}\beta_i\leq1\label{cond:sigmak},
\end{equation}
in which case 
\begin{equation}
	\sigma_j=1+\sum_{i=1}^{j-1}\beta_i\left(\max\left(1,\frac{1}{\beta_j}\right)-\max\left(1,\frac{1}{\beta_i}\right)\right)-\max\left(1,\frac{1}{\beta_j}\right).
\end{equation}
When $L=2$ this leads to 
\begin{equation}
	\sigma =
	\begin{cases}
		0 &\text{if }1<\beta_1<\beta_2,\\
		\beta_1-1 &\text{if }1-\beta_2<\beta_1<1,\,\beta_2>1,\\
		(\beta_1-1)/\beta_2 &\text{if }1-\beta_2<\beta_1<1,\,\beta_2<1,\\
		-1 &\text{if }\beta_1+\beta_2<1.	\label{eq:sigmak_2l}		
	\end{cases}
\end{equation} 
We present these results in Fig.~\ref{fig:SI_avgk}a. We can detect three regions: In region I, geometry in both layers is strong and the average mutual degree is constant. In region III ($\beta_2<1-\beta_1$) both layers have a very low geometric coupling. In this case, the average mutual degree decays behaves as if the two layers were uncorrelated. Region II represents the intermediate regime where the average mutual degree vanished in the thermodynamic limit but where decay is slow due to geometric effects. 

\subsection{Perfectly correlated layers, heterogeneous degree distribution}\label{sec:generalgammak}
Where in the previous section we completely ignored the popularity dimension and focused solely on similarity, here we begin with the opposite situation. We set $\beta=0$, such that the connection to the similarity space is completely lost. The generative model of the individual layers is now the soft configuration model, where the expected degrees can be fixed in expectation. We assume both layers share the same $\kappa$'s and that they are power-law distributed $\rho(\kappa)\propto\kappa^{-\gamma}$ with the same exponent and with the natural cut-off $\kappa_c\sim N^{1/(\gamma-1)}$~\cite{Boguna2004}. 

In this setting, let us first investigate the scaling of the average mutual degree. To this end we apply Eq.~\eqref{eq:M_generalh} to the SCM with perfect interlayers correlations, where $\boldsymbol{h}=\{\kappa,\kappa\}^T$, leading to 
\begin{equation}
	\langle \widetilde{k} \rangle = N\iint d\kappa'd\kappa'' \rho(\kappa')\rho(\kappa'') p(\kappa',\kappa'')^2.\label{eq:avgk_SCM}
\end{equation}
This equation can be solved analytically, leading to the following result for the average degree 
\begin{equation}
	\langle \widetilde k\rangle\sim
	\begin{cases}
		N^{2-\gamma}\ln N &\text{ if }2<\gamma<3\\
		N^{-1}&\text{ if }\gamma>3\label{eq:sigmak_SCM}
	\end{cases}
\end{equation}
A more heterogeneous network evidently leads to more edge overlap. This was to be expected, as the hidden degrees reintroduce correlations between the layers. A hub in one layer will also be one in the other, and there will be many shared connections. Note that even though these hubs increase the edge overlap, popularity alone is not able to provide us with a macroscopic amount of mutual edge. It is only similarity that can do so when both layers are sufficiently geometric.

Moving away from the $\beta=0$ limit we turn to numerical integration of the generalization of Eq.~\ref{eq:avgk_SCM}, given by 
\begin{equation}
	\langle \widetilde{k} \rangle = \frac{N}{\pi}\iint d\kappa'd\kappa''d\theta \rho(\kappa')\rho(\kappa'') p(\kappa',\kappa'',\theta,0)^2.\label{eq:avgk_S1}
\end{equation}
to study the mutual behavior. Eq.~\ref{eq:avgk_S1} was obtained by using the spherical symmetry of the system to place one of the nodes at the origin. In Fig.~\ref{fig:SI_avgk}b we plot the scaling exponent $\sigma_k$ as a function of $\beta$ for $\gamma\in[2.5,3.5,6]$. In the strongly geometric regime $\beta>1$, a constant mutual degree is obtained for all $\gamma$. The networks become ever sparser as $\beta$ decreases, finally stabilizing around the non-geometric $\beta=0$ results of Eq.~\eqref{eq:sigmac_SCM}. These results are shown in the inset. Note that the discrepancy for $\gamma=2.5$ is due to the influence of the $\ln N$ prefactor in this equation.

\subsection{General interlayer correlations}
\begin{figure*}
	\centering
	\includegraphics[width=\linewidth]{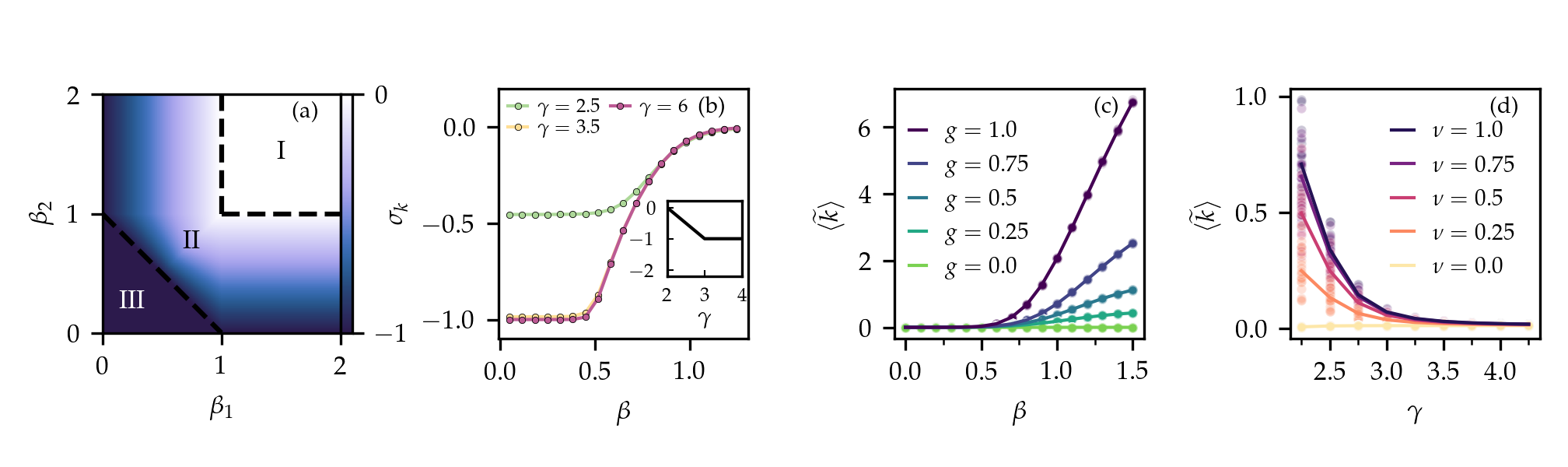}
	\vspace{-8mm}
	\caption{(a) The scaling exponent $\sigma_k$ of the average mutual degree $\km \sim N^{\sigma_k}$ as a function of the geometric couplings $\beta_1$ and $\beta_2$ of the constituent graphs for fully correlated layers and homogeneous degree distribution. The black dashed lines define 3 non-overlapping regions of the parameter space, based on how the average mutual degree relates to its single layer counterparts. 
	(b) The scaling exponents $\sigma_k$ as a function of $\beta=\beta_1=\beta_2$ for various power-law exponents $\gamma$. All results were obtained by numerical integration of Eq.~\ref{eq:avgk_S1} for $N\in[10^6,10^9]$. The inset shows the analytic result given in Eq.~\ref{eq:sigmak_SCM} for $\sigma_k$ as a function of $\gamma$ when $\beta=0$. 
	(c) The average mutual degree as a function of $\beta=\beta_1=\beta_2$ for various correlation strengths $g\in[0,1]$. In all cases the individual layers were generated with $N=32000$, $\gamma=50$, $\langle k\rangle = 20$ and $\nu=1$. 
	(d) The average mutual degree as a function of the degree distribution exponent $\gamma=\gamma_1=\gamma_2$. Various correlations strengths $\nu\in[0,1]$ are shown. The individual layers use $N=32000$, $\beta=0$, $\langle k\rangle =20$ and $g=1$. }
	\label{fig:SI_avgk}
\end{figure*}

We now generalize further and allow for general interlayer correlations. To this end we generate a set of GMMs with varying correlation parameters $g$ and $\nu$ using the method described in Appendix B in the main document. We then study the corresponding MGs. The results of this procedure are shown in Fig.~\ref{fig:SI_avgk}c,d. We see that lowering the correlations in the similarity and popularity dimensions leads to decreased edge overlap.

\newpage

\section{The mutual clustering coefficient}
In this section we study the scaling behaviors of the mutual clustering coefficient, given by the following equation 
\begin{equation}
	\overline{c}=\int d^L \boldsymbol{h}\, \mathbb{E}(c|\boldsymbol{h})\label{eq:c_marginalized_general},
\end{equation}
where
\begin{equation}
	\mathbb{E}(c|\boldsymbol{h}) = \frac{\iint d^L\boldsymbol{h'}d^L\boldsymbol{h''}\rho(\boldsymbol{h'})\rho(\boldsymbol{h''})\widetilde p(\boldsymbol{h},\boldsymbol{h'})\widetilde p(\boldsymbol{h},\boldsymbol{h''})\widetilde p(\boldsymbol{h'},\boldsymbol{h''})}{N^{-2}\mathbb{E}(k|\boldsymbol{h})^2}\label{eq:c_generalh},
\end{equation}
and where $\mathbb{E}(k|\boldsymbol{h})$ is the expected mutual degree of a node with hidden variables $\boldsymbol{h}$.

\subsection{Perfectly correlated layers, homogeneous degree distribution}
We start with the case of $L$ sparse, homogeneous, perfectly correlated layers with general $\{\beta_l\}_{l=1}^L$ and $\{\langle k\rangle^{(l)}\}_{l=1}^L$. In this setting, $\boldsymbol{h}=\{\theta^{(l)}\}_{l=1}^L$ and  $\rho(\theta^{(l)}|\theta^{(l-1)})=\delta(\theta^{(l)}-\theta^{(l-1)})$ $\forall l$, implying that Eq.~\eqref{eq:M_generalh} can be written as
\begin{equation}
	\langle \widetilde{c}\rangle=\left(\frac{N}{\langle \widetilde k\rangle\pi}\right)^2\int_0^\pi d\theta'\int_0^{\theta'} d\theta'' \prod_{l=1}^Lf_l(\theta',\theta'')\label{eq:c_S1_hom},
\end{equation}
where 
\begin{equation}
	f_l(\theta',\theta'')\equiv \frac{1}{1+(\zeta_l\theta')^{\beta_l}}\frac{1}{1+(\zeta_l\theta'')^{\beta_l}}\frac{1}{1+(\zeta_l(\theta'-\theta''))^{\beta_l}}.
\end{equation}
We notice that $\mathbb{E}(k|\boldsymbol{h})$ has reduced to $\langle \widetilde k\rangle$ because we are working in the homogeneous case. This allows us to focus now on the integral part of Eq.~\ref{eq:c_S1_hom}:
\begin{equation}
		\langle\widetilde{t}\rangle\equiv\int_0^\pi d\theta'\int_0^{\theta'}d\theta''\prod_{i=1}^{L}f(\theta',\theta''),
\end{equation}%

such that $\langle{\widetilde{c}}\rangle=\left(\frac{N}{\pi}\right)^2\langle\widetilde{t}\rangle/\langle\widetilde{k}\rangle^2$. We then once again assume that $\zeta_1\geq \zeta_2\geq ... \geq \zeta_m$, and split the $\theta'$ prime integral accordingly:
\begin{equation}
	\langle\widetilde t\rangle=\sum_{j=1}^{L+1}I_j,
\end{equation}
where
\begin{equation}
	I_j = \int_{\zeta_{j-1}^{-1}}^{\zeta_j^{-1}} d\theta'\int_0^{\theta'}d\theta''\prod_{i=1}^{L}f(\theta',\theta'').
\end{equation}
We can then find the upper and lower bound for each integral using the same logic as before. For the lower bound we have
{\begin{alignat}{6}
		I_j&\geq \int_{\zeta_{j-1}^{-1}}^{\zeta_j^{-1}} d\theta'\int_0^{\theta'}d\theta''\min_{\theta',\theta''}\left(\prod_{i=1}^{L}f(\theta',\theta'')\right)\notag\\[3mm]
		&=\frac{1}{2}\left(\zeta_j^{-2}-\zeta_{j-1}^{-2}\right)\prod_{i=1}^{L}f(\zeta_j^{-1},\zeta_j^{-1}/2)\sim \zeta_j^{-2}\prod_{i=1}^{j-1}\left(\frac{\zeta_i}{\zeta_j}\right)^{-3\beta_i},
\end{alignat}}%
where we again use that $\zeta_i/\zeta_j>1$ if $i<j-1$. For the upper bound we need to take some care. Naively, one starts as follows
{\begin{alignat}{6}
		I_j&\leq \int_{\zeta_{j-1}^{-1}}^{\zeta_j^{-1}} d\theta'\int_0^{\theta'}d\theta''\prod_{i=1}^{j-1}(\zeta_i^3\theta'\theta''(\theta'-\theta''))^{-\beta_i}\notag\\[3mm]
		&=\frac{\Gamma(1-a_j)^2}{(2-3a_j)\Gamma(2-2a_j)}\left(\zeta_j^{-2}\prod_{i=1}^{j-1}\left(\frac{\zeta_i}{\zeta_j}\right)^{-3\beta_i}-\zeta_{j-1}^{-2}\prod_{i=1}^{j-1}\left(\frac{\zeta_i}{\zeta_{j-1}}\right)^{-3\beta_i}\right)\label{eq:upperbound_t},
\end{alignat}}%
where we have defined $a_j=\sum_{i=1}^{j-1}\beta_i$ and where we again see that the second term is just the first for $j\rightarrow j-1$. However, this integral is only defined for $a_j<1$. We can thus only use this upper bound for $j$ such that $\sum_{i=1}^{j-1}\beta_i<1$. For higher $j$ we make a looser upper bound
\begin{alignat}{6}
	I_j&\leq \int_{\zeta_{j-1}^{-1}}^{\zeta_j^{-1}} d\theta'\int_0^{\theta'}d\theta''\prod_{i=1}^{\xi-1}(\zeta_i^3\theta'\theta''(\theta'-\theta''))^{-\beta_i},
\end{alignat}
where $\xi$ is chosen such that $\sum_{i=1}^{\xi-1}\beta_i<1$ and $\sum_{i=1}^{\xi}\beta_i\geq1$.  This integral can then be evaluated as before. However, the upper and lower bound do no longer match for $j>\xi$. This is not a problem as we will later show that other terms in the series will dominate even this upper bound, which means that the terms $j>\xi$ will always be sub-leading. Summarizing, we have found that 
\begin{equation}
	\sum_{j=1}^{\xi}I_j \sim \sum_{j=1}^{\xi} \zeta_j^{-2}\prod_{i=1}^{j-1}\left(\frac{\zeta_i}{\zeta_j}\right)^{-3\beta_i} = \sum_{j=1}^{\xi}N^{\sigma_j},
\end{equation}
and
{
	\begin{alignat}{6}
		I_j &\leq \frac{\Gamma(1-a_{\xi})^2}{(2-3a_\xi)\Gamma(2-2a_\xi)}\left(\zeta_j^{-2}\prod_{i=1}^{\xi-1}\left(\frac{\zeta_i}{\zeta_j}\right)^{-3\beta_i}-\zeta_{j-1}^{-2}\prod_{i=1}^{\xi-1}\left(\frac{\zeta_i}{\zeta_{j-1}}\right)^{-3\beta_i}\right)\notag\\[3mm]
		&\sim N^{\max(\tilde\sigma_j,\tilde\sigma_{j-1})}\quad\text{if }j>\xi.
\end{alignat}}%
First we find which of the $\sigma_j$ with $j\leq\xi$ dominates. We again look at two consecutive exponents:
\begin{equation}
	\sigma_{j}-\sigma_{j-1} = \left(\max\left(1,\frac{1}{\beta_j}\right)-\max\left(1,\frac{1}{\beta_{j-1}}\right)\right)\left(3\sum_{i=1}^{j-1}\beta_i-2\right).
\end{equation} 
Because $\zeta_{j-1}\geq \zeta_{j}$ we have that $\sigma_j\geq\sigma_{j-1}$ when $\sum_{i=1}^{j-1}\beta_i\leq\frac{2}{3}$. The dominant $\sigma_j$ is given by the $j$ that satisfies
\begin{equation}
	\frac{2}{3}-\beta_j\leq\sum_{i=1}^{j-1}\beta_i\leq\frac{2}{3}\label{cond:sigmat},
\end{equation}
in which case the scaling is given by
\begin{equation}
	\sigma_j=3\sum_{i=1}^{j-1}\beta_i\left(\max\left(1,\frac{1}{\beta_j}\right)-\max\left(1,\frac{1}{\beta_i}\right)\right)-2\max\left(1,\frac{1}{\beta_j}\right)\label{eq:scaling_t}.
\end{equation}
For $j>\xi$ we have 
\begin{equation}
	\tilde\sigma_{j}-\tilde\sigma_{j-1} = \left(\max\left(1,\frac{1}{\beta_j}\right)-\max\left(1,\frac{1}{\beta_{j-1}}\right)\right)\left(3\sum_{i=1}^{\xi-1}\beta_i-2\right).
\end{equation} 
Again, $\tilde\sigma_j\geq\tilde\sigma_{j-1}$ when $\sum_{i=1}^{\xi-1}\beta_i\leq\frac{2}{3}$. By definition of $\xi$ we also know that $\sum_{i=1}^{\xi}\beta_i\geq1$. This can only be true if $j=\xi$, which does not satisfy $j>\xi$. Thus, we know that $\tilde\sigma_{j-1}>\tilde\sigma_j$ for all $j>\xi$. The largest exponent is then given by the term $I_{\xi+1}\sim N^{\tilde\sigma_\xi}$. However, $\tilde\sigma_\xi=\sigma_\xi$ which is the scaling of $I_{\xi}$ for which we know the exact scaling, not just an upper bound. The upper bounds on $I_j$ for $j>\xi$, and therefore also the integrals themselves, are therefore all sub-leading to $I_{\xi}$.

We can thus conclude that Eq.~\eqref{eq:scaling_t} is the scaling of $\langle\widetilde{t}\rangle$. To obtain the scaling of the average local clustering coefficient $\langle\widetilde{c}\rangle=(N/\pi)^2\langle\widetilde{t}\rangle/\langle \widetilde{k}\rangle^2$ we use that the scaling exponents are related as follows
\begin{equation}
	\sigma_{i,j}^{(c)}=2+\sigma_i^{(t)}-2\sigma_j^{(k)}.
\end{equation}
Now, $i$ and $j$ are not necessarily the same. However, we can assume that $i\leq j$. This is because the condition given in Eq.~\eqref{cond:sigmak} is weaker than that given in Eq.~\eqref{cond:sigmat}. This then allows us to write $\sigma_{i,j}^{(c)}$ as 
\begin{alignat}{6}
	\sigma_{i,j}^{(c)} &= \sum_{l=1}^{i-1}\beta_l\left(3\max\left(1,\frac{1}{\beta_i}\right)-2\max\left(1,\frac{1}{\beta_j}\right)-\max\left(1,\frac{1}{\beta_l}\right)\right)\nonumber\\[3mm]
	&+2\sum_{l=i}^{j-1}\beta_l\left(\max\left(1,\frac{1}{\beta_l}\right)-\max\left(1,\frac{1}{\beta_j}\right)\right)\notag\\[3mm]
	&-2\left(\max\left(1,\frac{1}{\beta_i}\right)-\max\left(1,\frac{1}{\beta_j}\right)\right)\label{eq:scalingc}.
\end{alignat}
Note that the second sum starts at $l=i$, not $l=1$. A certain $\sigma_{i,j}^{(c)}$ dominates when 
\begin{equation}
	\frac{2}{3}-\beta_i\leq\sum_{l=1}^{i-1}\beta_l\leq\frac{2}{3}\quad\text{and}\quad1-\beta_j\leq\sum_{l=1}^{j-1}\beta_l\leq1\label{eq:genscalingexpsc}.
\end{equation}
In words this is equivalent to saying: The dominant scaling $\sigma_{i,j}^{(c)}$ is given when the sum of the smallest $i-1$ $\beta$'s is smaller than $2/3$ but where adding one more $\beta$ makes the sum larger than $2/3$. Simultaneously, the sum of the smallest $j-1$ $\beta$'s is smaller than $1$ but adding the $j$'th $\beta$ to the sum will make it larger than $1$.

In the case of two layers, Eq.~\ref{eq:genscalingexpsc} leads to the following scaling exponents
\begin{equation}
	\sigma^{(L=2)}_c(\beta_1,\beta_2) =
	\begin{cases}
		0 &\text{if }1<\beta_1<\beta_2,\\
		-2(1-\beta_1)^2/\beta_1 &\text{if }2/3<\beta_1<1,\,\beta_2>1\\
		-2(1-\beta_1)(\beta_2-\beta_1)/(\beta_1\beta_2) &\text{if }2/3\leq\beta_1\leq\beta_2\leq1,\\
		\beta_1-1 &\text{if }\beta_1\leq2/3,\,\beta_2\geq1,\\
		(\beta_1-\beta_2)/\beta_2 &\text{if }1-\beta_2\leq\beta_1<2/3,\,\beta_2<1,\\
		(\beta_2+3\beta_1-2)/\beta_2 &\text{if }2/3-\beta_2\leq\beta_1<2/3,\,\beta_2<1,\\	
		-2 &\text{if }\beta_1+\beta_2<2/3,\\			
	\end{cases}
\end{equation}
where we assume, without loss of generality, that $\beta_1\leq\beta_2$.

This also allows us to compare the multilayer case to the single layer scaling. In ref.~\cite{vanderKolk2022} it was shown that in the single layer case, the scaling exponent is given by 
\begin{equation}
	\sigma_c^{(L=1)}(\beta)=\begin{cases}
		0 &\text{if }\beta>1,\\
		2-2/\beta &\text{if }\beta>2/3,\\
		-1 &\text{else.}
	\end{cases}
\end{equation}
With this we can define the areas given in Fig.2a of the main text, based on whether the two layer scaling is layer or smaller than either or both of the single layer constituents. The results are given by
\begin{itemize}
	\item Area I: The two-layer decay is equal to both single layer scalings ($\sigma_c^{(L=2)}(\beta_1,\beta_2)=\sigma_c^{(L=1)}(\beta_1)=\sigma_c^{(L=1)}(\beta_2)=0$) if $\beta_1,\beta_2>1$.
	\item Area II: The two-layer decay is larger to the decay of layer $l=1$ but smaller than that of layer $l=2$.  ($\sigma_c^{(L=2)}(\beta_1,\beta_2)<\sigma_c^{(L=1)}(\beta_1)$, $\sigma_c^{(L=2)}(\beta_1,\beta_2)>\sigma_c^{(L=1)}(\beta_1)$) if $\beta_1<1$ if $\beta_2>1$, $\beta_1<1-\sqrt{(1-\beta_2)}$ if $1>\beta_2>8/9$, $\beta_1<3\beta_2-2$ if $8/9>\beta_2>3/4$, $\beta_1<\beta_2/3$ if $3/4>\beta_2>2/3$,  $\beta_1>2/3(1-\beta_2)$ if $2/3>\beta_2>0$. 
	\item Area III: The two-layer decay is smaller to the decay of layer $l=1$ but larger than that of layer $l=2$. ($\sigma_c^{(L=2)}(\beta_1,\beta_2)>\sigma_c^{(L=1)}(\beta_1)$, $\sigma_c^{(L=2)}(\beta_1,\beta_2)<\sigma_c^{(L=1)}(\beta_1)$) if $\beta_2<1$ if $\beta_1>1$, $\beta_2<1-\sqrt{(1-\beta_1)}$ if $1>\beta_1>8/9$, $\beta_2<3\beta_1-2$ if $8/9>\beta_1>3/4$, $\beta_2<\beta_1/3$ if $3/4>\beta_1>2/3$,  $\beta_2>2/3(1-\beta_1)$ if $2/3>\beta_1>0$. 
	\item Area V: The two layer decay is larger than that of both layers  ($\sigma_c^{(L=2)}(\beta_1,\beta_2)<\sigma_c^{(L=1)}(\beta_1)$, $\sigma_c^{(L=2)}(\beta_1,\beta_2)<\sigma_c^{(L=1)}(\beta_1)$) if $\beta_2<2/3(1-\beta_1)$ if $\beta_1>\beta_2$, $\beta_1<2/3(1-\beta_2)$ if $\beta_2>\beta_1$
	\item Area IV: The two layer decay is smaller than that of both layers ($\sigma_c^{(L=2)}(\beta_1,\beta_2)>\sigma_c^{(L=1)}(\beta_1)$, $\sigma_c^{(L=2)}(\beta_1,\beta_2)>\sigma_c^{(L=1)}(\beta_1)$) else.
\end{itemize}

In the case of general $L$, one notices that if $\beta_1=\beta_2=...=\beta_i=...=\beta_j$, all $\beta's$ that appear in Eq.~\eqref{eq:scalingc} are equal, which leads to $\sigma_{i,j}^{(c)}$ being zero. Thus, for a multiplex with $L$ layers, we have several manifolds in parameter space where the clustering is constant. These are given by 
\begin{alignat}{6}
	&1/2\leq\beta_1=\beta_2<\beta_3\leq...\leq\beta_L,\notag\\
	&1/3\leq\beta_1=\beta_2=\beta_3<\beta_4\leq...\leq\beta_L,\notag\\
	&\,\,\vdots\notag\\
	&1/L\leq\beta_1=\beta_2=...=\beta_L.
\end{alignat}
This final case where all couplings are equal can be understood by studying the effective connection probability of the mutual graph. It can be shown that for large networks, the connection probability of a single layer graph can be approximated by
\begin{equation}
	p'_{ij}=\min\left(1,(\zeta\Delta\theta_{ij})^{-\beta}\right)
\end{equation}
The mutual connection probability for a multiplex with $\beta_l=\beta\,\forall l$ and $\zeta_l=\zeta$ $\forall l$ is then just given by 
\begin{equation}
	\widetilde p'_{ij}=\prod_{l=1}^{L}\min\left(1,(\zeta\Delta\theta_{ij})^{-\beta}\right)=\min\left(1,(\zeta\Delta\theta_{ij})^{-L\beta}\right).
\end{equation}
If we now define $\widetilde\beta \equiv L\beta$, we observe that the effective mutual clustering coefficient is equivalent to the single layer version with re-scaled effective geometric coupling. In the $\mathcal{S}^1$, the transition point between finite and vanishing clustering lies at $\beta=1$. In the $L$-layer case, it will therefore lie at $\widetilde \beta=1$, or, equivalently, at $\beta_1=\beta_2=...=\beta_L=1/L$. 
\subsection{Perfectly correlated layers, heterogeneous degree distribution}\label{sec:generalgammac}
We now study the effect of degree heterogeneity. We first set $\beta=0$, such that we are working in the SCM. We assume both layers share the same $\kappa$'s and that they are power-law distributed $\rho(\kappa)\propto\kappa^{-\gamma}$ with the same exponent and with the natural cut-off $\kappa_c\sim N^{1/(\gamma-1)}$~\cite{Boguna2004}. 

In this setting, Eq.~\ref{eq:c_generalh} reduces to
\begin{equation}
	\mathbb{E}(c|\kappa)=\frac{\int d\kappa'd\kappa''\rho(\kappa')\rho(\kappa'') p(\kappa',\kappa'')^2p(\kappa,\kappa'')^2p(\kappa',\kappa)^2}{N^{-2}\mathbb{E}(k|\kappa)^2}\label{eq:clustering_SCM}.
\end{equation}
In order to obtain the average local clustering coefficient, one would need to marginalize once more over $\kappa$. However, as argued in Ref.~\cite{ColomerdeSimon2012}, $\mathbb{E}(c|\kappa)$ is a monotonously decreasing function, implying that one only needs to study $\mathbb{E}(c|\kappa)$ for some constant small $\kappa$ in order to obtain the dominant scaling. Following similar steps to the one taken in the aforementioned reference, it can be proven that 
\begin{equation}
	\overline{c}\sim\begin{cases}
		N^0&\text{ if }2<\gamma<3\\
		N^{-4\frac{\gamma-3}{\gamma-1}}&\text{ if }3<\gamma<5\\
		N^{-2}&\text{ if }5<\gamma\label{eq:sigmac_SCM}.
	\end{cases}
\end{equation}
Moving away from $\beta=0$ we numerically integrate the generalization of Eq.~\ref{eq:clustering_SCM} given by 
\begin{alignat}{6}
	\mathbb{E}(c|\kappa)=
	&\frac{\iiiint\mathrm{d}\kappa'\mathrm{d}\kappa''\mathrm{d}\theta'\mathrm{d}\theta''\rho(\kappa')\rho(\kappa'')p(\kappa,\kappa',\pi-|\pi-|\theta'||)^2p(\kappa,\kappa'',\pi-|\pi-|\theta''||)^2p(\kappa',\kappa'',\pi-|\pi-|\theta'-\theta''||)^2)}{\iint\mathrm{d}\kappa'\mathrm{d}\theta'\rho(\kappa')p(\kappa,\kappa',\pi-|\pi-|\theta'||)^2}.
\end{alignat}
Here, the final argument of the connection probabilities represents the distance between nodes along the circle. Note that we have once again made use of the spherical symmetry of our system, setting $\theta=0$. The results of this procedure are shown in Fig.~2b in the main text.

\newpage

\section{Randomized networks}

Degree preserving randomization, achieved by performing double edge swaps on networks, has been shown to be effective way to draw samples from the microcanonical configuration model~\cite{Cobb2003}. When applied to a geometric network, it effectively decouples the graph from its underlying metric space and is equivalent to setting the geometric coupling $\beta$ to zero. We use this procedure to separate the effects of the similarity and popularity dimension on the mutual average degree and mutual clustering coefficients observed in the real multiplexes in Fig.~1 in the main text. We perform $5\langle k\rangle N$ rewirings on both layers of the pairwise MCCs. We then construct the randomized MG by keeping only those edges that are present in both layers of the randomized pairwise MCCs.

\begin{figure*}[h]
	\centering
	\includegraphics[width=\linewidth]{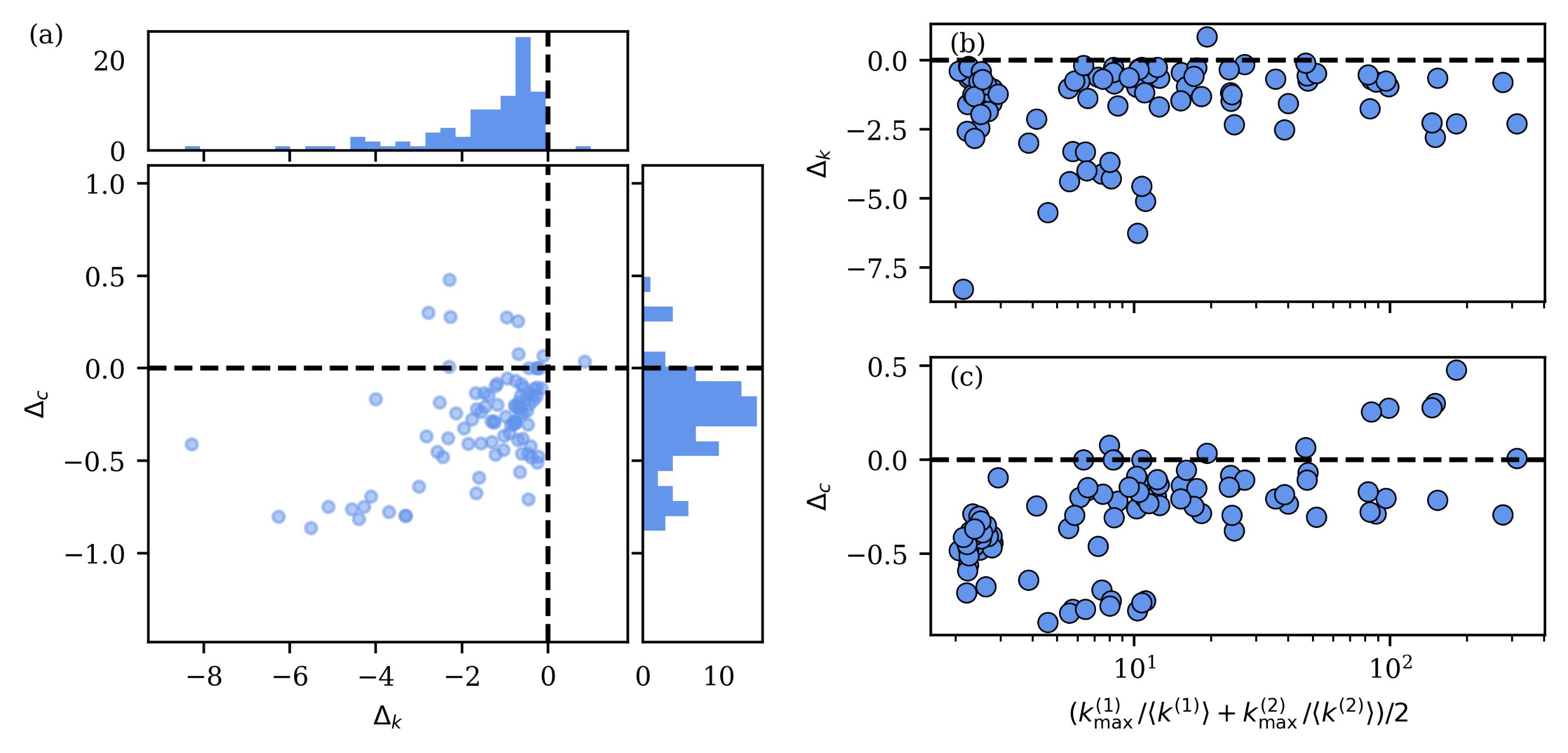}
	\vspace{-8mm}
	\caption{(a) The distribution of the differences between the average degree and average clustering coefficient of the mutual graph where the underlying networks have been randomized and the mutual graph of the original pairwise MCC. We define  $\Delta_k \equiv \langle \widetilde k^\prime\rangle - \langle \widetilde k\rangle$ and $\Delta_c \equiv \langle \widetilde c^\prime\rangle - \langle \widetilde c\rangle$, where the primed quantity refers to the randomized case. (b,c) The difference $\Delta_k$ and $\Delta_c$, respectively, as a function of the degree heterogeneity of the underlying graphs.}
	\label{fig:SI_randomized}
\end{figure*}

In Fig~\ref{fig:SI_randomized}a we show the distribution of the differences in average mutual degree $\Delta_k \equiv \langle \widetilde k^\prime\rangle - \langle \widetilde k\rangle$ and in the average mutual clustering coefficient $\Delta_c \equiv \langle \widetilde c^\prime\rangle - \langle \widetilde c\rangle$, where the primed quantities refers to the randomized case. We see that in most cases, both quantities decrease after randomization. This effect is strongest for the average degrees. This is in line with the analytical findings in Sec.~\ref{sec:generalgammak}, where we showed that $\sigma_k<0$ for $\beta_1=\beta_2=0$, irrespective of the degree heterogeneity of the underlying networks. In contrast, in Sec.~\ref{sec:generalgammac} it was shown that heterogeneous multiplexes with degree exponents $2<\gamma_1=\gamma_2<3$ scale with $\sigma_c=0$. We study the effect on degree heterogeneity on the quantities $\Delta_k$ and $\Delta_c$ in Figs.~\ref{fig:SI_randomized}b and c, respectively. As extracting the degree exponent from real data is not always possible we use the average relative size of the largest degree of the two underlying networks as a proxy for this quantity. We confirm that the effect of degree heterogeneity is stronger for the mutual clustering, where most pairwise MCCs with a small decrease or an increase in the mutual clustering of their corresponding MGs have, on average, a broad degree distribution. 

Finally, we note that most real networks studied in this work are relatively homogeneous and that, in most cases, clustering decays after randomization. These facts highlight the importance of geometry in these systems. 

\bibliography{bibliography}

%